# Edge AI without Compromise: Efficient, Versatile and Accurate Neurocomputing in Resistive Random-Access Memory

Weier Wan[1]*, Rajkumar Kubendran[2,5], Clemens Schaefer[4], S. Burc Eryilmaz[1], Wenqiang Zhang[3], Dabin Wu[3], Stephen Deiss[2], Priyanka Raina[1], He Qian[3], Bin Gao[3]*, Siddharth Joshi[4,2]*, Huaqiang Wu[3]*, H.-S. Philip Wong[1]*, Gert Cauwenberghs[2]*

[1] Stanford University, CA, USA;  [2] University of California San Diego, CA, USA;  [3] Tsinghua University, Beijing, China;  [4] University of Notre Dame, IN, USA;  [5] University of Pittsburgh, PA, USA

## Abstract

Realizing today's cloud-level artificial intelligence (AI) functionalities directly on devices distributed at the edge of the internet calls for edge hardware capable of processing multiple modalities of sensory data (e.g. video, audio) at unprecedented energy-efficiency. AI hardware architectures today cannot meet the demand due to a fundamental "memory wall": data movement between separate compute and memory units consumes large energy and incurs long latency[1]. Resistive random-access memory (RRAM) based compute-in-memory (CIM) architectures promise to bring orders of magnitude energy-efficiency improvement by performing computation directly within memory, using intrinsic physical properties of RRAM devices[2–7]. However, conventional approaches to CIM hardware design limit its functional flexibility necessary for processing diverse AI workloads, and must overcome hardware imperfections that degrade inference accuracy. Such trade-offs between efficiency, versatility and accuracy cannot be addressed by isolated improvements on any single level of the design. By co-optimizing across all hierarchies of the design from algorithms and architecture to circuits and devices, we present NeuRRAM - the first multimodal edge AI chip using RRAM CIM to simultaneously deliver a high degree of versatility in reconfiguring a single chip for diverse model architectures, record energy-efficiency 5× - 8× better than prior art across various computational bit-precisions, and inference accuracy comparable to software models with 4-bit weights on all measured standard AI benchmarks including accuracy of 99.0% on MNIST and 85.7% on CIFAR-10 image classification, 84.7% accuracy on Google speech command recognition, and a 70% reduction in image reconstruction error on a Bayesian image recovery task. This work paves a way towards building highly efficient and reconfigurable edge AI hardware platforms for the more demanding and heterogeneous AI applications of the future.

## Main

Compute-in-memory (CIM) architecture offers a pathway towards achieving brain-level information processing efficiency by eliminating expensive data movement between isolated compute and memory units in a conventional von Neumann architecture[2–7]. Resistive random access memory (RRAM)[8] is an emerging non-volatile memory that offers higher density, lower leakage and better analog programmability than conventional on-chip static random access memory (SRAM), making it an ideal candidate to implement large-scale and low-power CIM systems. Research in this area has demonstrated various AI applications by using fabricated resistive memory arrays as electronic synapses while using off-chip software/hardware to implement essential functionalities such as analog-to-digital conversion and neuron activations for a complete system[3,4,16,17,7,9–15]. More recent studies have demonstrated fully-integrated RRAM-CMOS chips and focused on techniques to improve energy-efficiency[18–28]. However, to date, there has not been a fully-integrated RRAM CIM chip that simultaneously demonstrates a broad cross-section of AI applications while achieving high energy-efficiency and software-comparable accuracy obtained entirely from hardware measurements. The challenge originates from the fundamental trade-offs between efficiency, versatility, and accuracy in hardware design. The analog nature of RRAM-CIM architecture makes it challenging to realize the same level of functional flexibility and computational accuracy as digital circuits. As a result, few CIM designs have implemented the reconfigurability necessary for processing diverse AI workloads, which is key to a broader adoption of the technology (e.g. the ability of graphics processing units[29] and tensor processing units[30] to execute diverse AI workloads contributes greatly to their adoption). Meanwhile, although various techniques have been proposed to mitigate the impacts of analog-related hardware imperfections on inference accuracy, the AI benchmark results reported in prior studies are often generated by performing software emulation based on characterized device data[4,7,11,17]. Such an approach often overestimates accuracies compared to fully hardware measured results due to incomplete modeling of hardware non-idealities. Obtaining software comparable inference results across broad AI applications remains challenging for RRAM CIM hardware.

To address these challenges, we present NeuRRAM - the first multimodal monolithically-integrated RRAM CIM hardware that simultaneously delivers a high degree of reconfigurability, record computational energy-efficiency, together with software comparable inference accuracy measured across various AI applications (Fig. 1). This is realized through innovations across the full stack of the design: (1) at the device level, 3 million RRAM devices with high analog programmability and uniformity are monolithically integrated with CMOS circuits; (2) at the circuit level, a voltage-mode neuron circuit supports variable computation bit-precision and diverse neuron activation functions while performing analog-to-digital conversion at low power consumption and compact area footprint; (3) at the architecture level, a bi-directional transposable neurosynaptic array (TNSA) architecture is developed to enable reconfigurability in dataflow directions with minimal area and energy overheads; (4) at the system level, the chip is partitioned into 48 cores that perform inference in parallel and can be selectively powered on/off to allow flexible control of computation granularity to achieve high energy efficiency; (5) finally at the algorithm level, hardware-algorithm co-optimization techniques including model-driven hardware calibration and hardware-driven model training and fine-tuning are developed to mitigate the impact of hardware non-idealities on inference accuracy. We report fully hardware measured benchmark results for a broad range of AI inference tasks including image classification, voice recognition and image recovery, implemented with diverse AI models including convolutional neural networks (CNNs)[31], long short-term memory (LSTM)[32] models and probabilistic graphical models[33] (Fig. 1e). The chip is measured to consistently achieve the lowest energy-delay-product (EDP), an often used metric of performance and energy-efficiency, among RRAM CIM hardware, while it operates over a range of configurations to suit various AI benchmarks (Fig. 1d).

**Reconfigurable RRAM CIM architecture for diverse AI workloads**

To support a broad range of AI workloads, the NeuRRAM chip implements a highly flexible CIM chip architecture. The chip consists of 48 CIM cores that can perform computation in parallel. Each individual core can be selectively turned off through power-gating when not actively used, while the model weights are retained by the non-volatile RRAM devices. At the central of each core is a novel Transposable Neurosynaptic Array (TNSA) architecture that offers flexible control of dataflow directions, which is crucial for enabling diverse model architectures with different dataflow patterns. For instance, in CNNs that are commonly applied to vision-related tasks, data flows in a single direction through layers to generate data representations at different abstraction levels; in LSTMs that are used to process temporal data such as audio signals, data travel recurrently through the same layer for multiple time-steps; in probabilistic graphical models such as restricted Boltzmann machine (RBM), probabilistic sampling is performed back-and-forth between layers until the network converges to a high-probability state. Besides inference, the error back-propagation during gradient-descent training of multiple AI models requires reversing the direction of dataflow through the network. However, conventional RRAM CIM architectures are limited to perform MVM in a single direction by hardwiring rows and columns of the RRAM crossbar array to dedicated circuits on the periphery to drive inputs and measure outputs. Some studies implement reconfigurable data-flow directions by adding extra hardware, which incurs substantial energy, latency and area penalties (Extended Data Fig. 2): executing bi-directional (forward and backward) data-flow requires either duplicating power and area-hungry analog-to-digital converters (ADCs) at both ends of the RRAM array[22,34] or dedicating large area outside of the array to routing both rows and columns to a shared data converter[26]; the recurrent connections require writing the outputs to a buffer memory outside of the RRAM array, and reading them back for the next time step computation[35].

The TNSA architecture (Fig. 2b, c) realizes dynamic data-flow reconfigurability with little overhead. While in conventional designs, CMOS peripheral circuits such as ADCs connect only at one end of the RRAM array, the TNSA architecture physically interleaves the RRAM weights and the CMOS neuron circuits that implement ADCs and activation functions, and interconnects them along the length of both rows and columns (Fig. 2b-d). The physical layout of TNSA consists of 16×16 of such interleaved corelets that are connected by shared bit-lines (BLs) and word-lines (WLs) along horizontal direction and source-lines (SLs) along vertical direction. Each corelet encloses 16×16 RRAM devices and one neuron circuit. To distribute the 256 neurons across the TNSA to 256 rows (BL) and 256 columns (SL), we connect the neuron within the corelet ($i, j$) to both the ($16i+j$)-th BL and the ($16j+i$)-th SL via a pair of switches. As a result, each row or each column connects uniquely to a single neuron, while doing so without duplicating neurons at both ends of the array, thus saving area and energy. Moreover, a neuron uses its BL and SL switches for both its input and output: it not only receives the analog matrix-vector-multiplication (MVM) output coming from either BL or SL through the switches, but also sends the converted digital results to peripheral registers via the same switches. By configuring which switch to use during the input and output stages of the neuron, we can realize multiple MVM data-flow directions. Fig. 2e shows the forward, backward and recurrent MVMs enabled by the TNSA. To implement forward MVM (BL-to-SL), during the input stage, input pulses are applied to the BLs through the BL drivers, get weighted by the RRAMs, and enter the neuron through its SL switch; during the output stage, the neuron sends the converted digital outputs to bottom SL registers through its SL switch; to implement recurrent MVM (BL-to-BL), the neuron instead receives input through its SL switch and sends the digital output back to the BL registers through its BL switch.

To realize a balance between the flexibility of computation granularity and area- and energy- efficiency, we partition the chip area into 48 CIM cores, with each core containing a 256×256 RRAM cells and 256 CMOS neuron circuits. While a larger core size would amortize area and power consumption of peripheral circuits (e.g., digital-to-analog and analog-to-digital converters) across more RRAM devices, it is too inflexible and inefficient for implementing small, irregular, or sparse weight matrices. Conversely, a larger multiplicity of smaller cores provides finer control over computation granularity, making the design more versatile for a wide range of model architectures, at the expense of reduced memory density and higher energy consumed by peripheral circuits. Moreover, the multi-core design supports a broad selection of weight mapping strategies, which allows us to exploit both model-parallelism and data-parallelism to maximize computation throughput (Fig. 2a). Using a CNN as an example, to maximize data-parallelism, we duplicate the weights of the most computationally intensive layers (early convolutional layers) to multiple cores for parallel inference on multiple data; to maximize model-parallelism, we map different convolutional layers to different cores and perform parallel inference in a pipelined fashion. Meanwhile, we divide the layers whose weight dimensions exceed RRAM array size into multiple segments and assign them to multiple cores for parallel execution. A more detailed description of the weight mapping strategies is provided in the Methods.

**Efficient and reconfigurable voltage-mode neuron circuit**

The NeuRRAM chip achieves 5× to 8× lower energy-delay-product (EDP) and 20× to 61× higher peak computational throughput at various MVM input and output bit-precisions than previously reported RRAM CIM chips[18–28] (Fig. 1d), where the energy and delay are measured for performing an MVM with a 1024×1024 weight matrix. Key to the energy-efficiency and throughput improvement is a novel in-memory MVM output sensing scheme. The conventional approach is to use voltage as input, and measure current as the results based on Ohm's law[18–22,24,26,27] (Fig. 2g). Such a current-mode sensing scheme has limitations on energy-efficiency, throughput, and versatility: (1) simultaneously applying voltage pulses to multiple rows and columns leads to large array current, consuming large power; (2) sinking the large current while clamping the voltage requires large transistors at the transimpedance amplifier (TIA), consuming large area; and (3) MVM within different AI models produces drastically different output dynamic ranges that span several orders of magnitude (Fig. 2i). Optimizing the ADC across such a wide dynamic range is fundamentally difficult. To equalize the dynamic range and reduce the array current, designs typically limit the number of input wires to activate in a single cycle, thus limiting throughput.

NeuRRAM simultaneously improves energy-efficiency, throughput, and versatility by virtue of a highly reconfigurable neuron circuit implementing a voltage-mode sensing scheme. The neuron performs analog to digital conversion of the MVM outputs by directly sensing the settled open-circuit voltage on the BL or SL line capacitance[36], obviating the need for power and area hungry TIA to sense current while clamping voltage (Fig. 2h). Voltage inputs are driven on the BLs and voltage outputs are sensed on the SLs, or vice versa, depending on the desired direction of MVM. WLs are activated to start the MVM operation, activating all RRAM conductances. The voltage on the output line settles to the weighted average of the voltages driven on the input lines, where the weights are the RRAM conductances. Upon deactivating the WLs, the output is sampled by transferring the charge on the output line to the neuron sampling capacitor ($C_{sample}$ in Extended Data Fig. 4). The neuron then accumulates this charge onto an integration capacitor ($C_{integ}$). The analog-to-digital conversion of the integrated charge starts when all the sampling and integration cycles during the MVM input phase complete (see details in Methods). Such voltage-mode sensing turns off the RRAM array immediately when voltage on output wires settles, before the ADC process starts, thus requiring a shorter array activation time and lower energy consumption than existing implementations of current-mode sensing[36]. Moreover, the weight normalization due to the conductance weighting in the voltage output (Fig. 2i) results in an automatic output dynamic range normalization for different weight matrices, making the MVM self-adaptable to different AI models while not sacrificing throughput. To eliminate the normalization factor from the final results, we pre-compute its value from the weight matrix and multiply it back to the digital outputs from the ADC.

Our voltage-mode neuron further supports reconfigurability in computation bit-precision and activation functions. It can perform MVM with 1- to 6-bit inputs and 1- to 8-bit outputs, and implement commonly used activation functions such as Rectified Linear Unit (ReLU), sigmoid, and tanh functions as a part of the analog-to-digital conversion process. It also supports probabilistic sampling for stochastic activation functions by injecting pseudo-random noise generated by a linear-feedback-shift-register (LFSR) block into the neuron integrator. All the neuron circuit operations are performed by dynamically configuring a single amplifier in the neuron as either an integrator or a comparator during different phases of operations, as detailed in the Methods. This results in a more compact design than other work that merges ADC and neuron activation functions within the same module[23,28]. While most existing CIM designs use time-multiplexed ADCs for multiple rows/columns to amortize ADC area, the compactness of our neuron circuit (1.2 μm² each) allows us to dedicate a neuron for each pair of BL and SL, and tightly interleave the neuron with RRAM devices within the TNSA architecture as can be seen in Extended Data Fig. 10d.

**Hardware-algorithm co-optimization enables software comparable inference accuracy**

The innovations across the chip architecture and circuit layers bring superior efficiency and reconfigurability to NeuRRAM. To complete the story, we must ensure that the computational accuracy can be preserved under various circuit and device non-idealities[4,6,37]. We developed a set of hardware-algorithm co-optimization techniques that allow NeuRRAM to deliver software-comparable accuracy across diverse AI applications. Importantly, all the AI benchmark results presented in the paper are obtained entirely from hardware measurements. While prior efforts have reported benchmark results using a mixture of hardware characterization and software simulation, e.g., emulate the array-level MVM process in software using measured single device characteristics[4,7,11,17], such an approach often fails to model the complete set of non-idealities existing in realistic hardware. As shown in Fig. 3a, they include (i) IR drop on input wires, (ii) on RRAM array drivers, and (iii) on crossbar wires, (iv) limited RRAM programming resolution, (v) RRAM conductance relaxation, (vi) capacitive coupling from simultaneously switching array wires, and (vii) limited ADC resolution and dynamic range. Our experiments show that omitting certain non-idealities in simulation leads to over-optimistic prediction of inference accuracy when compared to actual hardware measurement results. For example, the 3rd and the 4th bars in Fig. 3e show a 2.32% accuracy difference between simulation and measurement for CIFAR-10 classification[38], while the simulation accounts for only (v) RRAM conductance relaxation and (vii) limited ADC resolution, which are what previous studies most often modelled[7,17].

Our hardware-algorithm co-optimization approach includes three main techniques: (1) model-driven chip calibration, (2) noise-resilient NN training and analog weight programming, and (3) chip-in-the-loop progressive model fine-tuning. Model-driven chip calibration uses the real model weights and input data to optimize the chip operating conditions such as input voltage pulse amplitude, and finds any ADC offsets for subsequent cancellation during inference. Ideally, the MVM output voltage dynamic range should fully utilize the ADC input swing to minimize the discretization error. However, MVM output dynamic range varies with network layers (Fig. 3b). As a result, for each individual layer, we use a subset of training-set data to that layer to search for the optimal operating conditions. Extended Data Fig. 5 shows that different input data distributions lead to vastly different output distributions for a layer. Therefore, it is crucial to perform the optimization using the training-set data that can mostly closely emulate the input data distribution we will see at test time. Together with the weight normalization effect from the voltage-mode neuron circuit, the model-driven chip calibration technique enables the chip to flexibly adapt to different AI models without sacrificing ADC accuracy.

Stochastic non-idealities such as RRAM conductance relaxation degrade signal-to-noise ratio (SNR) of computation, leading to inference accuracy drop. Prior work obtained higher SNR by limiting each RRAM cell to store only a single bit, and encoding higher precision weights using multiple cells[18,21,24,27,39]. Such an approach lowers the weight memory density and energy efficiency. Accompanying that approach, the neural network is trained with weights quantized to the corresponding bit-precisions. In contrast, we utilize the intrinsic analog programmability of RRAM[40] to maximize weight memory density while maintaining accuracy. Instead of training with quantized weights, which is equivalent to injecting uniformly distributed noise into weights, we train the model with high-precision floating-point weights while injecting noise with distribution extracted from characterization of actual RRAM devices. A previous study proposed to inject Gaussian noise into weights to improve the network's noise resiliency[17]. Our technique can be more generally applied to different types of resistive memories whose noise does not necessarily follow a Gaussian distribution. Fig. 3e shows that the technique significantly improves the model's immunity to noise, from a CIFAR-10 classification accuracy of 25.34% without noise injection to 85.99% with noise injection. After the training, we program the non-quantized weights to RRAM analog conductances using an iterative write-verify technique, described in the Methods. This technique enables NeuRRAM to achieve an inference accuracy comparable to software models with 4-bit weights across various applications (Fig. 1e). Meanwhile, each weight is encoded by the differential conductance of 2 RRAM cells, which improves weight memory density by 2× compared to previous studies that require one RRAM cell per bit.

By applying the above two techniques, we already can measure inference accuracy comparable to or better than software models with 4-bit weights on many tasks, including Google speech command recognition[41], MNIST[42] image recovery, and MNIST classification (Fig. 1e). For deeper neural network models that are more sensitive to hardware non-idealities such as ResNet-20[43], the measured accuracy on CIFAR-10 classification (83.67%) is still 3.36% lower than that of a 4-bit-weight software model (87.03%). The accuracy loss can be attributed to those non-idealities that are both difficult to model accurately in software during neural network training and cannot be compensated through hardware calibration. For instance, during multi-core parallel operations, large currents flowing through input wires and RRAM array drivers leads to large IR drops (non-ideality (i) and (ii) in Fig. 3a), causing significant accuracy degradation.

To further improve accuracy for deep neural networks, we introduce a novel chip-in-the-loop progressive fine-tuning technique. Chip-in-the-loop training allows to circumvent fabrication induced non-idealities by measuring training error directly on the chip[44]. Prior work showed that fine-tuning the final layers using the back-propagated gradients calculated from hardware measured outputs and re-programming the fine-tuned RRAM weights helped improve accuracy[7]. We find this technique to be of limited effectiveness in countering the non-idealities that have a non-linear impact on MVM outputs, such as IR drops. Weight reprogramming also consumes extra time and energy. Our chip-in-the-loop progressive fine-tuning overcomes non-linear model errors by exploiting the intrinsic non-linear universal approximation capacity of the deep neural network[45], and furthermore eliminates the need for weight reprogramming. Fig. 3d illustrates the fine-tuning procedure. We progressively program the weights one layer at a time onto the chip. After programming a layer, we perform inference using the training set data on the chip up to that layer, and use the measured outputs to fine-tune the remaining layers that are still training in software. In the next time step, we program and measure the next layer on the chip. We repeat this process until all the layers are programmed. During the process, the non-idealities of the programmed layers can be progressively compensated by the remaining layers through training. Fig. 3f shows the efficacy of the progressive fine-tuning technique. From left to right, each data point represents a new layer programmed onto the chip. The accuracy at each layer is evaluated by using the chip-measured outputs from that layer as inputs to the remaining layers in software. The cumulative CIFAR-10 test set inference accuracy is improved by 1.99% using this technique. Extended Data Fig. 7a further illustrates the extent to which fine-tuning recovers the training set accuracy loss at each layer, demonstrating the effectiveness of the approach in bridging the accuracy gap between software and hardware measurements.

Using the techniques described above, we implement a variety of AI models on the NeuRRAM chip, and achieve inference accuracy comparable to software models with 4-bit weights across all the measured AI benchmark tasks. Fig. 1e and Extended Data Fig. 7b show that we achieve a 0.98% error rate on MNIST handwritten digit recognition using a 7-layer CNN, a 14.34% error rate on CIFAR-10 object classification using ResNet-20, a 15.34% error rate on Google speech command recognition using a 4-cell LSTM, and a 70% reduction of L2 image reconstruction error compared to the original noisy images on MNIST image recovery using an RBM. Fig. 4 summarizes the key features of each demonstrated model. Most of the essential neural network layers and operations are implemented fully on the chip, including all the convolutional, fully-connected and recurrent layers, and neuron activation functions, batch-normalization, and stochastic sampling process. Each of the models is implemented by allocating the weights to multiple cores on a single NeuRRAM chip. The implementation details are described in the Methods. Fundamentally, Each of the selected models represents a general class of AI algorithms: CNNs represent the class of feed-forward deep neural networks commonly used for computer vision related tasks; LSTM represents the recurrent neural networks often used for processing time series such as audio signals; and RBM represents probabilistic graphical models that require bi-directional dataflow direction to perform probabilistic sampling among a large set of random variables. These results demonstrate the versatility of the NeuRRAM architecture and the wide applicability of the hardware-algorithm co-optimization techniques.

The NeuRRAM chip implements reconfigurability across the entire hierarchy of the design, from a multi-core architecture offering flexible computation granularity, to a transposable neurosynaptic array structure enabling dynamically reconfigurable dataflow direction, to a neuron circuit implementing diverse activation functions and bit-precisions, while doing all this at a record computational energy-efficiency among resistive memory CIM hardware. We expect the energy-efficiency (EDP) to improve by another 2 to 3 orders of magnitude when scaling the design from 130-nm to 7-nm CMOS and RRAM technologies while adopting a circuit architecture most suited for each technology node (detailed in Methods). The ability to process a diverse collection of AI models efficiently and accurately opens the door towards broader adoption of the CIM technology. Fundamentally, NeuRRAM signifies the importance of cross-layer co-optimization between device, circuit, architecture, and algorithm for ameliorating the trade-offs between efficiency, versatility, and accuracy in designing next-generation edge AI hardware. The techniques presented in this work can be more generally applied to any non-volatile resistive memory technology such as phase change memory[46] (PCM), conductive bridge random access memory[47] (CBRAM), ferroelectric field-effect transistor[48] (FeFET), and electro-chemical random access memory[49] (ECRAM). As resistive memory continues to scale towards offering tera-bits of on-chip memory[50], such a cross-layer co-optimization approach will equip CIM hardware on the edge with sufficient performance, efficiency, and versatility to perform complex AI tasks that can only be done on the cloud today.

## Methods

**Core block diagram and operating modes**

Fig. 2b and Extended Data Fig. 1 show the block diagram of a single compute-in-memory core. To support versatile MVM directions, most of the design is symmetrical in the row (BLs and WLs) and column (SLs) directions. The row and column register files store the inputs and outputs of MVMs, and can be written from both the external interface to the core via either an SPI or a random-access interface, and the neurons internal to the core. The SL peripheral circuits contain a linear feedback shift register (LFSR) block used to generate pseudo-random sequences used for probabilistic sampling. It is implemented by two LFSR chains propagating in opposite directions. The registers of the two chains are XORed to generate spatially uncorrelated random numbers[51]. The controller block receives commands and generates control waveforms to the BL/WL/SL peripheral logic and to the neurons. It contains a delay-line based pulse generator with tunable pulse width from 1 ns to 10 ns. It also implements clock-gating and power-gating logic used to turn off the core in idle mode. Each WL, BL and SL of the TNSA is driven by a driver consisting of multiple pass gates that supply different voltages. Based on the values stored in the register files and the control signals issued by the controller, the WL/BL/SL logic decides the state of each pass gate.

The core has mainly three operating modes: weight-programming mode, neuron testing mode, and MVM mode (Extended Data Fig. 1). In the weight-programming mode, individual RRAM cells are selected for read and write. To select a single cell, the registers at the corresponding row and column are programmed to "1" through random-access with the help of the row and column decoder, while the other registers are reset to "0". The WL/BL/SL logic turns on the corresponding driver pass gates to apply a SET/RESET/READ voltage on the selected cell. In the neuron testing mode, the WLs are kept at GND. Neurons receive inputs directly from BL or SL drivers through their BL or SL switch, bypassing RRAM devices. This allows us to characterize the neurons independently from the RRAM array. In the MVM mode, each input BL/SL is driven to $V_{ref}$ - $V_{read}$, $V_{ref}$ + $V_{read}$, or $V_{ref}$ depending on the registers' value at that row/column. If the MVM is in the BL-to-SL direction, we activate the WLs that are within the input vector length while keeping the rest at GND; if the MVM is in the SL-to-BL direction, we activate all the WLs. After neurons finish ADC, the pass gates from BLs/SLs to the registers are turned on to allow neuron state readout.

**Device fabrication**

RRAM arrays in NeuRRAM are in a one-transistor-one-resistor (1T1R) configuration, where each RRAM device is stacked on top of and connects in series with a selector NMOS transistor that cuts off the sneak path and provides current compliance during RRAM programming and reading. The selector NMOS, CMOS peripheral circuits, and the bottom four back-end-of-line (BEOL) interconnect metal layers are fabricated in a standard 130-nm foundry process. Due to the higher voltage required for RRAM forming and programming, the selector NMOS and the peripheral circuits that directly interface with RRAM arrays use thick-oxide I/O transistors rated for 5V operation. All the other CMOS circuits in neurons, digital logic, registers, etc. use core transistors rated for 1.8V operations.

The RRAM device is sandwiched between metal-4 and metal-5 layers shown in Fig. 2f. After the foundry completes the fabrication of CMOS and the bottom four metal layers, we use a laboratory process to finish the fabrication of RRAM devices and metal-5 interconnect, and top metal pad and passivation layers. The RRAM device stack consists of a TiN bottom electrode layer, a HfO$x$ switching layer, a TaO$x$ thermal enhancement layer[52], and a TiN top electrode layer. They are deposited sequentially, followed by a lithography step to pattern the lateral structure of the device array.

**RRAM write-verify programming and conductance relaxation**

Each neural network weight is encoded by the differential conductance between two RRAM cells on adjacent rows along the same column. The first RRAM cell encodes positive weight, and is programmed to low conductance state ($g_{min}$) if the weight is negative; the second cell encodes negative weight, and is programmed to $g_{min}$ if the weight is positive. Mathematically, the conductances of the two cells are $max(g_{max}\frac{W}{w_{max}}, g_{min})$ and $max(-g_{max}\frac{W}{w_{max}}, g_{min})$ respectively, where $g_{max}$ and $g_{min}$ are the maximum and minimum conductance of the RRAMs; $w_{max}$ is the maximum absolute value of weights; and $W$ is the unquantized high-precision weight.

To program an RRAM cell to its target conductance, we use an incremental-pulse write-verify technique[40]. Extended Data Figs. 3b, 3c illustrate the procedure: we start by measuring the initial conductance of the cell. If the value is below the target conductance, we apply a weak SET pulse aiming to slightly increase the cell conductance. Then we read the cell again. If the value is still below the target, we apply another SET pulse with amplitude incremented by a small amount. We repeat such SET-read cycles until the cell conductance is within an acceptance range to the target value or overshoots to the other side of the target. In the latter case we reverse the pulse polarity to RESET, and repeat the same procedure as with SET. During the SET/RESET pulse train, the cell conductance is likely to bounce up and down multiple times until eventually it enters the acceptance range or reaches a timeout limit. There are a few trade-offs in selecting programming conditions: (1) a smaller acceptance range and a higher time-out limit improve programming precision, but require a longer time; (2) a higher $g_{max}$ improves signal-to-noise ratio, but leads to higher energy consumption and more programming failures for cells that cannot reach high conductance. In our experiments, we set the initial SET pulse voltage to be 1.2V and RESET to be 1.5V, both with an increment of 0.1V and pulse width of 1μs. The acceptance range is ±1uS to the target conductance. The timeout limit is 30 SET-RESET polarity reversals. We used $g_{min}$ = 1 μS for all the models, and $g_{max}$ = 40 μS for CNNs and 30 μS for LSTMs and RBMs. With such settings, 99% of the RRAM cells can be programmed to the acceptance range within the timeout limit. On average each cell requires 8.52 SET/RESET pulses.

Besides the longer programming time, another reason to not use an overly small write-verify acceptance range is RRAM conductance relaxation. RRAM conductance drifts over time after programming. Most of the drift happens within a short time window (< 1s) immediately following the programming, after which the drift becomes much slower. The abrupt initial drift is called "conductance relaxation" in the literature[37]. Its statistics follow a Gaussian distribution at all conductance states except when the conductance is close to $g_{min}$. Extended Data Fig. 3d shows the conductance relaxation measured across the whole $g_{min}$ to $g_{max}$ conductance range. We found that the loss of programming precision due to conductance relaxation is much higher than that caused by the write-verify acceptance range. The measured standard deviation of the relaxation is ~2.8 μS, which is close to 10% of $g_{max}$. To mitigate the relaxation, we use an iterative programming technique. We iterate over the RRAM array for multiple times. In each iteration, we measure all the cells and re-program those whose conductance has drifted outside the acceptance range. The tail distribution becomes narrower with more programming iterations. After 3 iterations, the standard deviation becomes ~2 µS, a 29% decrease compared to the initial value. We use 3 iterations in all our neural network demonstrations, and perform inference at least 30 mins after the programming such that the measured inference accuracy would account for such conductance relaxation effects. By combining the iterative programming with our hardware-aware model training approach, the impact of relaxation can be largely mitigated.

**Implementation of MVM with multi-bit inputs and outputs**

The neuron and the peripheral circuits support MVM at configurable input and output bit-precisions. An MVM operation consists of an initialization phase, an input phase, and an output phase. Extended Data Fig. 4 illustrates the neuron circuit operation. During the initialization phase (Extended Data Fig. 4a), all BLs and SLs are precharged to $V_{ref}$. The sampling capacitors $C_{sample}$ of the neurons are also precharged to $V_{ref}$, while the integration capacitors $C_{integ}$ are discharged.

During the input phase (Extended Data Fig. 4b, e), each input wire (either BL or SL depending on MVM direction) is driven to one of three voltage levels: $V_{ref}$ - $V_{read}$, $V_{ref}$, $V_{ref}$ + $V_{read}$ through three pass gates, configured by a 2-bit register and a one-hot decoder for each input wire. With the differential row weight encoding, the input wires of a differential pair are driven to the opposite voltage with respect to $V_{ref}$. i.e. when input is 0, both wires are driven to $V_{ref}$; when input is +1, the two wires are driven to $V_{ref}$ + $V_{read}$ and $V_{ref}$ - $V_{read}$; and when input is -1, to $V_{ref}$ - $V_{read}$ and $V_{ref}$ + $V_{read}$. The multi-bit inputs are realized by accumulating multiple such (-1, 0, +1) pulses as will be explained later. We then turn on all the WLs that have inputs, and the output wires begin to approach their final steady-state voltages $V_j = \frac{\sum_i V_i G_{ij}}{\sum_i G_{ij}}$. When the voltages of the output wires settle, we turn off the WLs to shut down the current flow. We then sample the charge from the output wire capacitance to $C_{sample}$, and integrate the charge onto $C_{integ}$. Since $C_{sample}$ is much smaller than the output wire capacitance, such sampling and integration operations can be repeated for multiple cycles to increase the amount of integrated charge. We use this capability to implement MVM with multi-bit input. For example, when the input vectors are 4-bit signed integers with 1 sign-bit and 3 magnitude-bits, we send three voltage pulses to input wires. The first pulse corresponds to the 1st (most significant) magnitude-bit. When the bit is 0, we send a "0" pulse; when the bit is 1, we send a "+1" pulse if the sign-bit is 1, and "-1" pulse if the sign-bit is 0. We then sample and integrate from the output wires for 4 cycles such that the integrated charge quadruples. Similarly, for the 2nd or the 3rd magnitude-bit, we apply pulses to the input wires, but this time we sample and integrate for 2 cycles and 1 cycle, respectively. In general, for n-bit signed integer inputs, we need in total (n-1) input pulses and $2^{(n-1)}$-1 sampling and integration cycles.

Finally, during the output phase (Extended Data Figs. 5c, d, f), we convert the integrated charge into a multi-bit digital output. First, to generate the sign-bit, we disconnect the feedback loop of the amplifier to turn the integrator into a comparator (Extended Data Fig. 4c). We drive the right side of $C_{integ}$ to $V_{ref}$. If the integrated charge is positive, the comparator output will be driven to GND, and to VDD otherwise. The output is then inverted, latched, and sent to the BL or SL via the neuron BL/SL switch before being written into the peripheral BL/SL registers. To generate the magnitude-bits, we perform a "charge decrement" operation to the integrated charge (Extended Data Fig. 4d): if the sign-bit is 1, we charge the $C_{sample}$ to $V_{decr-}$ = $V_{ref}$ - $V_{decr}$ via the outer loop, and to $V_{decr+}$ = $V_{ref}$ + $V_{decr}$ otherwise. The charge is then transferred to $C_{integ}$ to cancel part of the charge integrated during the input phase. The updated voltage of $C_{integ}$ is again compared against 0V, and the comparator output is recorded. We repeat this process and count the number of charge decrement steps until the cumulative decremental charge completely cancels the initial charge, which causes the comparator output to flip polarity. The counter outputs constitute the magnitude-bits of the neuron output. We set the maximum number of charge decrement steps $N_{max}$ to be 128, so the output can be at maximum 8-bit (1 sign-bit + 7 magnitude-bits). To improve latency, we perform an early stop before $N_{max}$ is reached if the comparator outputs of all the neurons flip polarity. To configure the output bit-precision, we can tune $V_{decr}$ to change the sensing granularity and tune $N_{max}$ to change the total voltage range that the charge decrement process will cover.

We implement various commonly used neuron activation functions by modifying the operations of the charge decrement step. To implement Rectified Linear Unit (ReLU), we only perform charge decrement to resolve the magnitude-bits if the sign-bit is 1. This way we save energy by eliminating the charge decrement operations for neurons with negative integrated voltage. To implement sigmoid or tanh functions, instead of incrementing the counter every charge decrement step, we increase the number of steps between counter increments with the counter value. For instance, the counter value is initially incremented every step; when the counter value reaches 35, it begins to be incremented every 2 steps; when it reaches 40, every 3 steps; when reaches 43, every 4 steps, etc. This way a piecewise linear approximation of the sigmoid/tanh curve can be produced. For sigmoid function, we add the maximum number of steps to the tanh neuron output and divide it by twice that number to normalize the final output within the range [0, 1].

To summarize, both the configurable MVM input and output bit-precisions and various neuron activation functions are implemented using different combinations of the four basic operations: sampling, integration, comparison and charge decrement. Importantly, all the four operations are realized by a single amplifier configured in different feedback modes. As a result the design realizes versatility and compactness at the same time.

**Noise-resilient NN training**

For all the AI models that we implement, we start by training a baseline model using the noise resilient NN training technique: we inject noise into all the fully-connected and convolutional layers whose weights will be implemented by RRAM arrays during the forward-pass of NN training. The distribution of the injected noise is obtained from chip characterization: using the iterative write-verify technique, the RRAM conductance relaxation is characterized to have absolute value of mean < 1 μS ($g_{min}$) at all conductance states. The highest standard deviation is 3.87 μS, about 10% of the maximum conductance 40 μS, found at ~12 μS conductance state (Extended Data Fig. 3d). Therefore, when simulating model performance on software, we inject a Gaussian noise with a zero mean and a standard deviation equal to 10% of the maximum absolute value of weights of a layer.

To select the models that achieve the best inference accuracy at the 10% noise level, we train models with different levels of noise injection from 0% to 30%. We find that injecting a higher noise during training than testing improves models' noise resiliency. Extended Data Fig. 6a, b show that the best test-time accuracy in the presence of 10% weight noise is obtained with 20% and 15% training-time noise injection for CIFAR-10 image classification and Google voice command classification, respectively. For RBM, Extended Data Fig. 6c shows that the noise injection improves the inference performance even in the absence noise, possibly due to the probabilistic nature of RBM. The model trained with the highest noise injection (25%) achieves the lowest image reconstruction error.

Extended Data Fig. 6d shows the effect of noise injection on weight distribution. Without noise injection, the weights have a Gaussian distribution. The neural network outputs heavily depend on a small fraction of large weights, and thus become vulnerable to noise injection. With noise injection, the weights distribute more uniformly, making the model more noise-resilient.

To efficiently implement the models on NeuRRAM, inputs to all convolutional and fully-connected layers are quantized to 4-bit or below. The input bit-precision of all the models is summarized in Fig. 4a. We perform the quantized training using the Parameterized Clipping Activation (PACT) technique[53].

**Chip-in-the-loop progressive fine-tuning**

During the progressive chip-in-the-loop fine-tuning, we use the chip-measured intermediate outputs from a layer to fine-tune the weights of the remaining layers. Importantly, to fairly evaluate the efficacy of the technique, we do not use the test-set data (for either training or selecting checkpoint) during the entire process of fine-tuning. To avoid over-fitting to a small-fraction of data, measurements should be performed on the entire training-set data. We reduce the learning rate to 1/100 of the initial learning rate used for training the baseline model, and finetune for 30 epochs. The same weight noise injection and input quantization are applied during the fine-tuning. The cumulative CIFAR-10 test set inference accuracy is improved by 1.99% using the technique. Extended Data Fig. 7a further shows the training set accuracy loss at each layer is partially recovered by the fine-tuning, demonstrating the effectiveness of the approach in bridging the accuracy gap between software and hardware measurements.

**Implementations of CNNs, LSTMs and RBMs**

We use CNN models for the CIFAR-10 and MNIST image classification tasks. The CIFAR-10 dataset consists of 50,000 training images and 10,000 testing images belonging to 10 object classes. We perform image classification using the ResNet-20 convolutional neural network architecture[43] that contains 21 convolutional layers and 1 fully-connected layer (Fig. 4b), with batch-normalizations and ReLU activations in between the layers. The model is trained using the Keras framework. We quantize the input of all convolutional and fully-connected layers to a 3-bit unsigned fixed point format except for the first convolutional layer, where we quantize the input image to 4-bit because the inference accuracy is more sensitive to the input layer quantization. For the MNIST hand-written digits classification, we use a 7-layer CNN consisting of 6 convolutional layers and 1 fully-connected layer, and using max-pooling in between layers to down-sample feature map sizes. The inputs to all the layers, including the input image, are quantized to a 3-bit unsigned fixed-point format.

All the parameters of the CNNs are implemented on a single NeuRRAM chip including those of the convolutional layers, the fully-connected layers, and the batch-normalization. Fig. 4d illustrates the process to map a convolutional layer on a chip. To implement the weights of a 4D convolutional layer with dimension $H$ (height), $W$ (width), $I$ (# input channels), $O$ (# output channels) on 2D RRAM arrays, we flatten the first three dimensions into a 1D-vector, and append the bias term of each output channel to each vector. If the range of the bias values is $B$ times of the weight range, we evenly divide the bias values and implement them using $B$ rows. Further, we merge the batch normalization parameters with weights and biases after training (Fig. 4c), and program the merged $W'$ and $b'$ onto RRAM arrays such that no explicit batch normalization needs to be performed during inference. Under the differential-row weight mapping scheme, parameters of a convolutional layer are converted into a conductance matrix of size $(2(HWI+B),$ $O$). If the conductance matrix fits into a single core, an input vector is applied to 2($HWI+B$) rows and broadcasted to $O$ columns in a single cycle. $HWIO$ multiply-accumulate operations are performed in parallel. Most ResNet-20 convolutional layers have a conductance matrix height of 2($HWI+B$) that is greater than the RRAM array length of 256. We therefore split them vertically into multiple segments, and map the segments either onto different cores that are accessed in parallel, or onto different columns within a core that are accessed sequentially. The details of weight mapping strategies are described in the next section.

The Google speech command dataset consists of 65,000 one-second-long audio recordings of voice commands, such as "yes", "up", "on", "stop", etc.., spoken by thousands of different people. The commands are categorized into 12 classes. Fig. 4d illustrates the model architecture. We use the mel-frequency cepstral coefficient (MFCC) encoding approach to encode every 40 ms piece of audio into a length-40 vector. With a hop length of 20 ms, we have a time series of 50 steps for each one-second recording. We build a model that contains four parallel LSTM cells. Each cell has a hidden state of length 112. The final classification is based on summation of outputs from the four cells. Compared to a single-cell model, the 4-cell model reduces the classification error (of an unquantized model) from 10.13% to 9.28% by leveraging additional cores on the NeuRRAM chip. Within a cell, in each time step, we compute the values of four LSTM gates (input, activation, forget, output) based on the inputs from the current step and hidden states from the previous step. We then perform element-wise operations between the four gates to compute the new hidden state value. The final logit outputs are calculated based on the hidden states of the final time step. Each LSTM cell has 3 weight matrices that are implemented on the chip: an input-to-hidden-state matrix with size 40×448, a hidden-state-to-hidden-state matrix with size 112×448, and a hidden-state-to-logits matrix with size 112×12. The element-wise operations are implemented on the FPGA. The model is trained using the PyTorch framework. The inputs to all the MVMs are quantized to 4-bit signed fixed-point formats. All the remaining operations are quantized to 8-bit.

Restricted Boltzmann Machine (RBM) is a type of generative probabilistic graphical model. Instead of being trained to perform discriminative tasks such as classification, it learns the statistical structure of the data itself. Fig. 4e shows the architecture of our image-recovery RBM. The model consists of fully-connected 794 visible neurons, corresponding to 784 image pixels plus 10 one-hot encoded class labels, and 120 hidden neurons. We train the RBM using the contrastive divergence learning procedure in software. During inference, we send binarized images with partially corrupted or blocked pixels to the model running on a NeuRRAM chip. The model then performs back-and-forth MVMs and Gibbs sampling between visible and hidden neurons for 10 cycles. In each cycle, neurons sample binary states $h$ and $v$ from the MVM outputs based on the probability distributions: $p(h_j = 1 \mid v) = \sigma(b_j + \sum_i v_i w_{ij})$ and $p(v_i = 1 \mid h) = \sigma(a_i + \sum_j h_j w_{ij})$, where $a_i$ is a bias for hidden neurons ($h$) and $b_j$ is a bias for visible neurons ($v$). After sampling, we reset the uncorrupted pixels (visible neurons) to the original pixel values. The final inference performance is evaluated by computing the average L2 reconstruction error between the original image and the recovered image. Extended Data Fig. 8 shows some examples of the measured image recovery.

When mapping the 794×120 weight matrix to multiple cores of the chip, we try to make the MVM output dynamic range of each core relatively consistent such that the recovery performance will not overly rely on the computational accuracy of any single core. To achieve this, we assign adjacent pixels (visible neurons) to different cores such that every core sees a down-sampled version of the whole image (Fig. 4g). Utilizing the bi-directional MVM functionality of the transposable neurosynaptic array (TNSA), the visible-to-hidden neuron MVM is performed from SL to BL direction in each core; the hidden-to-visible neuron MVM is performed from BL to SL direction.

**Weight mapping strategy onto multiple CIM cores**

To implement an AI model on a NeuRRAM chip, we convert the weights, biases, and other relevant parameters (e.g. batch-normalization) of each model layer into a single two-dimensional conductance matrix as described in the previous section. If the height or the width of a matrix exceed the RRAM array size of a single CIM core (256×256), we split the matrix into multiple smaller conductance matrices, each with maximum height and width of 256.

We consider three factors when mapping these conductance matrices onto the 48 cores: resource utilization, computational load balancing and IR drop. The top priority is to ensure that all conductance matrices of a model are mapped onto different addresses within a single chip such that no re-programming is needed during inference. If the total number of conductance matrices does not exceed 48, we can map each matrix onto a single core (case 1 in Fig. 2a) or multiple cores. There are two scenarios when we map a single matrix onto multiple cores: (1) when a model has different computational intensities, defined as the amount of computation per weights, for different layers, e.g., CNNs often have higher computational intensity for earlier layers due to larger feature map dimensions, we duplicate the more computationally intensive matrices to multiple cores and operate them in parallel to increase throughput and balance the computational loads across the layers (case 2 in Fig. 2a); (2) Some models have "wide" conductance matrices (output dimension > 128), such as our image-recovery RBM. If mapping the entire matrix onto a single core, the IR drop on the array drivers due to the large total RRAM conductance along the rows can deteriorate inference accuracy. When there are spare cores, we can split the matrix vertically into multiple segments and map them onto different cores to mitigate the IR drop (case 6 in Fig. 2a).

On the other hand, if a model has more than 48 conductance matrices, we need to merge some matrices so that they can fit onto a single chip. The smaller matrices are merged diagonally such that they can be accessed in parallel (case 3 in Fig. 2a); The bigger matrices are merged horizontally and accessed sequentially due to shared rows (case 4 in Fig. 2a). When selecting the matrices to merge, we want to avoid the matrices that belong to the same two categories described in the previous paragraph: (1) those that have high computational intensity (e.g. early layers of ResNet-20) to minimize impact on throughput; and (2) those with "wide" output dimension (e.g. late layers of ResNet-20 have large number of output channels) to avoid large IR drop. For instance, in our ResNet-20 implementation, among a total of 61 conductance matrices (see Fig. 4a: 1 from input layer, 12 from block 1, 17 from block 2, 28 from block 3, 2 from shortcut layers, and 1 from final dense layer), we map each matrix in block 1 and block 3 onto a single CIM core, and merge the remaining 21 matrices into 8 larger matrices to occupy the 8 remaining CIM cores. Although the segments that share the same rows within a merged matrix need to be accessed sequentially, they do not affect the overall throughput of the ResNet-20, which is limited by the more computationally intensive matrices in block 1.

**Test system implementation**

Extended Data Fig. 9a shows the hardware test system for the NeuRRAM chip. The NeuRRAM chip (left) is configured by, receives inputs from and sends outputs to a Xilinx Spartan-6 field-programmable gate array (FPGA) that sits on an Opal Kelly integrated FPGA board (right). The FPGA communicates with the PC via a USB 3.0 module. The test board also houses voltage DACs that provide various bias voltages required by RRAM programming and MVM, and ADCs to measure RRAM conductance during the write-verify programming. The power of the entire board is supplied by a standard "cannon-style" DC power connector and integrated switching regulators on the Opal Kelly board such that no external lab equipment is needed for the chip operation.

To enable fast implementation of various machine learning applications on the NeuRRAM chip, we developed a software toolchain that provides Python based application programming interfaces (APIs) at various levels. The low level APIs provide access to basic operations of each chip module such as RRAM read and write and neuron analog-to-digital conversion; the middle level APIs include essential operations required for implementing neural network layers such as the multi-core parallel MVMs with configurable bit-precision and RRAM write-verify programming; the high level APIs integrate various middle-level modules to provide complete implementations of neural network layers, such as weight mapping and batch inference of convolutional and fully-connected layers. The software toolchain aims to allow software developers who are not familiar with the NeuRRAM chip design to deploy their machine learning models on the NeuRRAM chip.

**Power and throughput measurements**

To characterize MVM energy efficiency at various input and output bit-precisions, we measure the power consumption and latency of MVM input and output stages separately. The total energy consumption and the total time are the sum of input and output stages because the two stages are performed independently as described in the above sections. As a result, we can easily obtain the energy-efficiency of any combinations of input and output bit-precisions.

To measure the input stage energy efficiency, we generate a 256×256 random weight matrix with Gaussian distribution, split it into 2 segments, each with dimension 128×256, and program the two segments to two cores using the differential-row weight mapping. We measure the power consumption and latency for performing 10 million MVMs, or equivalently 655 billion multiply-accumulate (MAC) operations. The comparison with prior work shown in Fig. 1d uses the same workload as benchmark.

Extended Data Fig. 10a shows how the energy per operation changes with the input bit-precision. The inputs are in the signed integer format, where the first bit represents sign, and the rest of the bits represent magnitude. 1-bit (binary) and 2-bit (ternary) shows similar energy because each input wire is driven to 1 of 3 voltage levels. Binary input is therefore just a special case for ternary input. The output stage energy efficiency is measured by performing parallel analog-to-digital conversion using 256 neurons in a core. Similar to the inputs, the outputs are also represented in the signed integer format. Extended Data Fig. 10b shows that the measured energy per conversion grows exponentially with output bit-precision due to the exponentially increasing charge-decrement cycles as shown in Extended Data Fig. 4f.

Extended Data Fig. 10c shows the power consumption breakdown. A large fraction of energy is spent in switching on and off the WLs that connect to select transistors of RRAM devices. These transistors use thick-oxide I/O transistors in order to withstand high-voltage during RRAM forming and programming. They are sized large enough (width = 1μm, length = 500 nm) to provide sufficient current for RRAM programming. As a result, they require high operating voltages and add large capacitance to the WLs, both contributing to high power consumption ($P = f\ C\ V^2$). This energy is expected to go down significantly if RRAMs can be written by a lower voltage and have a lower conductance state, and if a smaller transistor with better drivability can be used.

The fraction of power spent in neuron operation and digital control grows with bit precision, which can be expected from the neuron operation: Extended Data Figs. 5e and 5f show that the number of sampling and integration cycles during input phase, and the number of comparison and charge decrement cycles during output phase both grow exponentially with bit-precision, whereas the number of WLs switching and MVM input pulses grow only linearly with input bit precision.

Finally, Extended Data Fig. 10d shows the peak throughput while performing both input and output stages of MVM. The output stage is assumed to use 2-bit higher precision than inputs to account for the additional bit-precision required for partial-sum accumulations. The required partial sum bit-precision for the voltage-mode sensing implemented by NeuRRAM is much lower than that required by the conventional current-mode sensing, because all the 256 input wires can be activated in a single cycle, leading to a reduced number of partial sum accumulation steps. The simultaneous activation of more wires also leads to a higher computational throughput. Compared to a 2Mb RRAM CIM macro using current-mode sensing implemented in 22nm[27], NeuRRAM implemented in 130nm achieves a 20× to 61× improvement in peak computational throughput at various bit-precisions. To quantify a combined measure of energy-efficiency and performance, energy-delay-product (EDP) is commonly used as a figure of merit in comparing designs than throughput-power efficiency as tera-operations per second per watt (TOPS/W, reciprocal of energy per operation) because it captures the time-to-solution aspect in addition to energy consumption. Nevertheless, we include the TOPS/W measurement results in Extended Data Fig. 10e.

**Projection of NeuRRAM energy-efficiency with technology scaling**

The current NeuRRAM chip is fabricated using a 130 nm CMOS technology. We expect the energy-efficiency to improve with the technology scaling. Importantly, isolated scaling of CMOS transistors and interconnects is not sufficient for the overall energy-efficiency improvement. RRAM device characteristics must be optimized jointly with CMOS. The current RRAM array density under one-transistor-one-resistor (1T1R) configuration is limited not by the fabrication process, but by the RRAM write current and voltage. The current NeuRRAM chip uses large thick-oxide I/O transistors as the "T" in order to withstand > 4V RRAM forming voltage and provide enough write current. Only if we lower both the forming voltage and write current can we obtain higher density and therefore lower parasitic capacitance for improved energy-efficiency.

Assuming RRAM devices at newer technology node can be programmed at a logic-compatible voltage level, and the required write current can be reduced such that the size of the connecting transistor keeps shrinking, the EDP improvements will come from (1) lower operating voltage and (2) smaller wire and transistor capacitance, i.e., $Energy \propto CV^2$, $Delay \propto CV/I$. At 7 nm for instance, we expect the WL switching energy (Extended Data Fig. 10c) to reduce by ~22.4×, including 2.6× from WL voltage scaling (1.3V → 0.8V), and 8.5× from WL length/capacitance scaling (same as minimum metal pitch scaling 340 nm → 40 nm). Peripheral circuit energy will reduce by at least ~ 5× from the sole VDD scaling (1.8V → 0.8V). The energy consumed by the MVM pulses and charge transfer process is independent of the range of RRAM conductance, since power consumption and settling time of the RRAM array scale with the same conductance factor that cancels in their product. Specifically the energy per RRAM multiply-accumulate (MAC) is $E_{MAC} = C_{par} \, var(V_{in})$, limited only by parasitic capacitance per unit RRAM cell $C_{par}$, and variance in driven input voltage $var(V_{in})$. Therefore, the MVM energy consumption will reduce by approximately ~34×, including 4× from read voltage scaling (0.5V → 0.25V), and 8.5× from smaller parasitic capacitance. Overall, conservatively, we expect an energy consumption reduction of ~8× when scaling the design from 130-nm to 7-nm. In terms of the latency, the current design is limited by the settling time of the neuron amplifier. At more advanced technology nodes, one could adopt a different architecture not based on voltage integration to achieve a higher speed. For instance, a Flash ADC using dynamic comparator can improve the latency for doing a 256×256 matrix vector multiplication with 4-bit output from 2.1 µs in the current NeuRRAM chip to 22 ns at 7-nm, giving a 95× improvement. Therefore, we expect the overall EDP to improve by ~760× when scaling the design from 130-nm to 7-nm technology and adopting circuit architecture most suited for each technology node.

## Data availability
The datasets used for benchmarks are publicly available[38,41,42]. Other data that support the findings of this study are available from the corresponding author upon reasonable request.

## Code availability
The NeuRRAM chip software toolchain described in Methods and the codes used to perform noise-resilient model training and chip-in-the-loop progressive model fine-tuning are available from the corresponding author upon reasonable request.

## Author Contributions
W.W., R.K., S.B.E., S.J., H.S.W. and G.C. designed the NeuRRAM chip architecture and circuits; W.W., S.B.E., W.Z. and D.W. implemented physical layout of the chip; W.Z., H.Q., B.G. and H.W. contributed the RRAM device fabrication and integration with CMOS; W.W., R.K., S.D. and G.C. developed the test system; W.W. developed the software toolchain, implemented the AI models on the chip and conducted all chip measurements; W.W., C.S. and S.J. worked on the development of AI models; W.W., R.K., C.S., P.R., S.J., H.P.W, G.C. contributed to the experiment design and analysis and interpretation of the measurements; B.G., S.J., H.W., H.S.P. and G.C. supervised the project. All authors contributed to the writing and editing of the manuscript.

## Competing Interests
The authors declare no competing interests.

## Materials & Correspondence
Correspondence to Weier Wan (weierwan@stanford.edu), Bin Gao (gaob1@tsinghua.edu.cn), Siddharth Joshi (sjoshi2@nd.edu), Huaqiang Wu (wuhq@tsinghua.edu.cn), H.-S. Philip Wong (hspwong@stanford.edu), and Gert Cauwenberghs (gert@ucsd.edu).

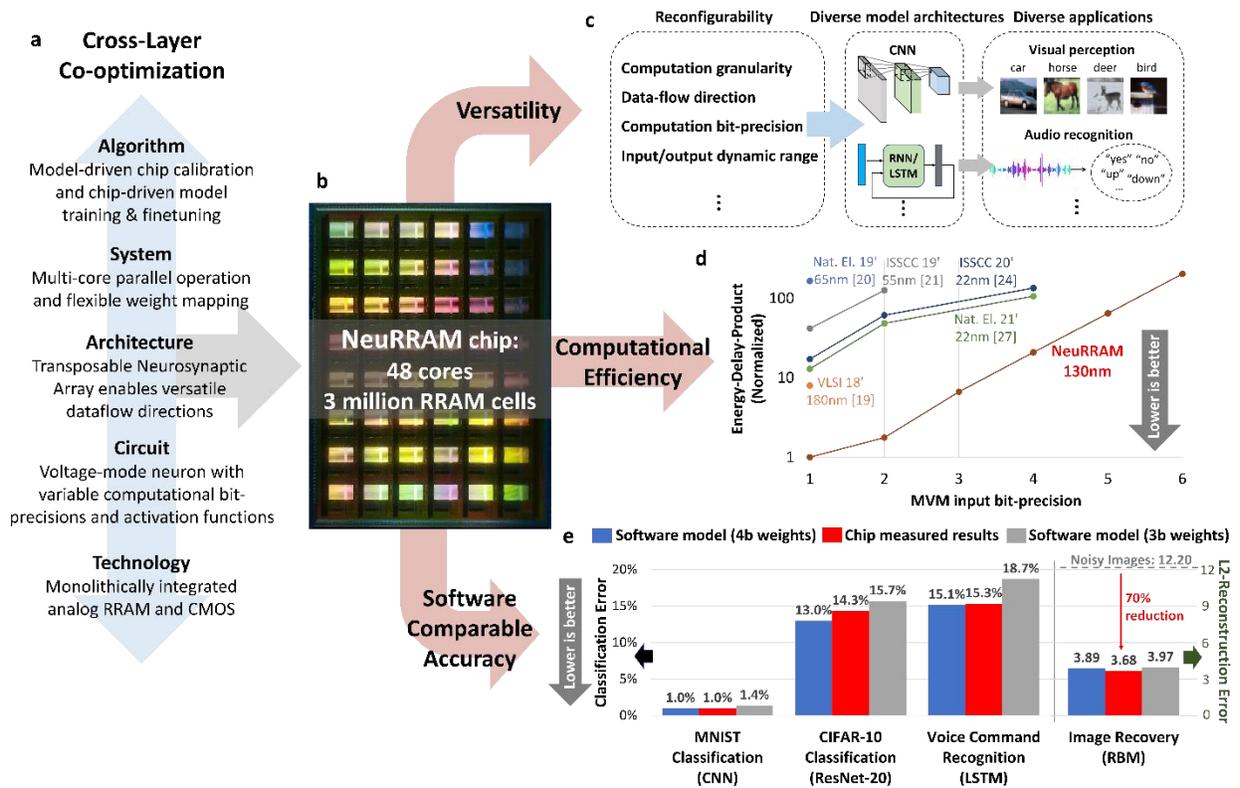

**Fig.1 | Design methodology and main contributions of the NeuRRAM chip. a**, Cross-layer co-optimizations across the full stack of the design enable NeuRRAM to simultaneously deliver high versatility, computational efficiency and software-comparable inference accuracy. **b**, Micrograph of the NeuRRAM chip. **c**, Reconfigurability in various aspects of the design enables NeuRRAM to implement diverse AI models for a wide variety of applications. **d**, Comparison of energy-delay-product (EDP) efficiency and performance metric among recent resistive memory compute-in-memory hardware. **e,** Fully hardware measured inference accuracy from NeuRRAM is comparable to software models across various applications.

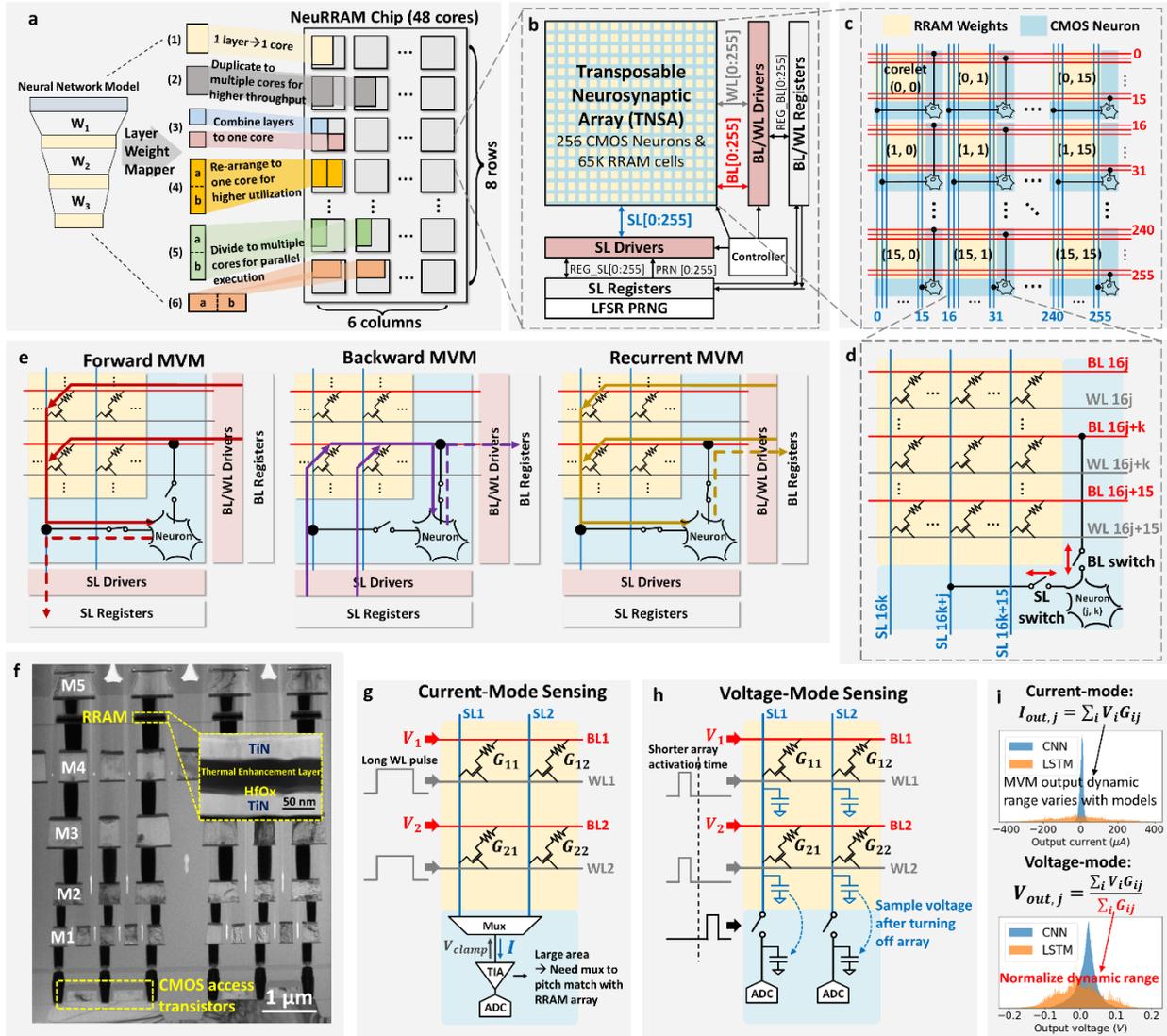

**Fig.2 | Design of the NeuRRAM chip. a**, Multi-core architecture of the NeuRRAM chip, and various ways to map neural network layers onto cores depending on layer weight shape, throughput and utilization requirements. **b**, Block diagram of a single compute-in-memory core. **c**, Architecture of the Transposable Neurosynaptic Array (TNSA) interleaving RRAM weights and CMOS neurons. It consists of 16×16 RRAM-CMOS corelets connected by shared horizontal bit-lines (BL) and word-lines (WL) and vertical source-lines (SL). **d,** Architecture of a TNSA corelet. Each corelet contains 16×16 RRAMs and 1 neuron. The neuron connects to both a BL and a SL via BL and SL switches used for both neuron's input and output. **e**, The TNSA can be dynamically configured for matrix-vector-multiplication (MVM) in forward, backward or recurrent directions. **f**, A cross-sectional transmission electron microscopy (TEM) image showing the layer stack of the monolithically integrated RRAM and CMOS. **g**, Conventional current-mode sensing scheme for in-memory matrix-vector multiplication. **h,** Voltage-mode sensing employed by NeuRRAM. Energy is saved by shortening array (WL) activation time. **i**, MVM output distribution from a CNN layer and from an LSTM layer (weights normalized to the same range.) Voltage-mode sensing normalizes wide variation in output dynamic range.

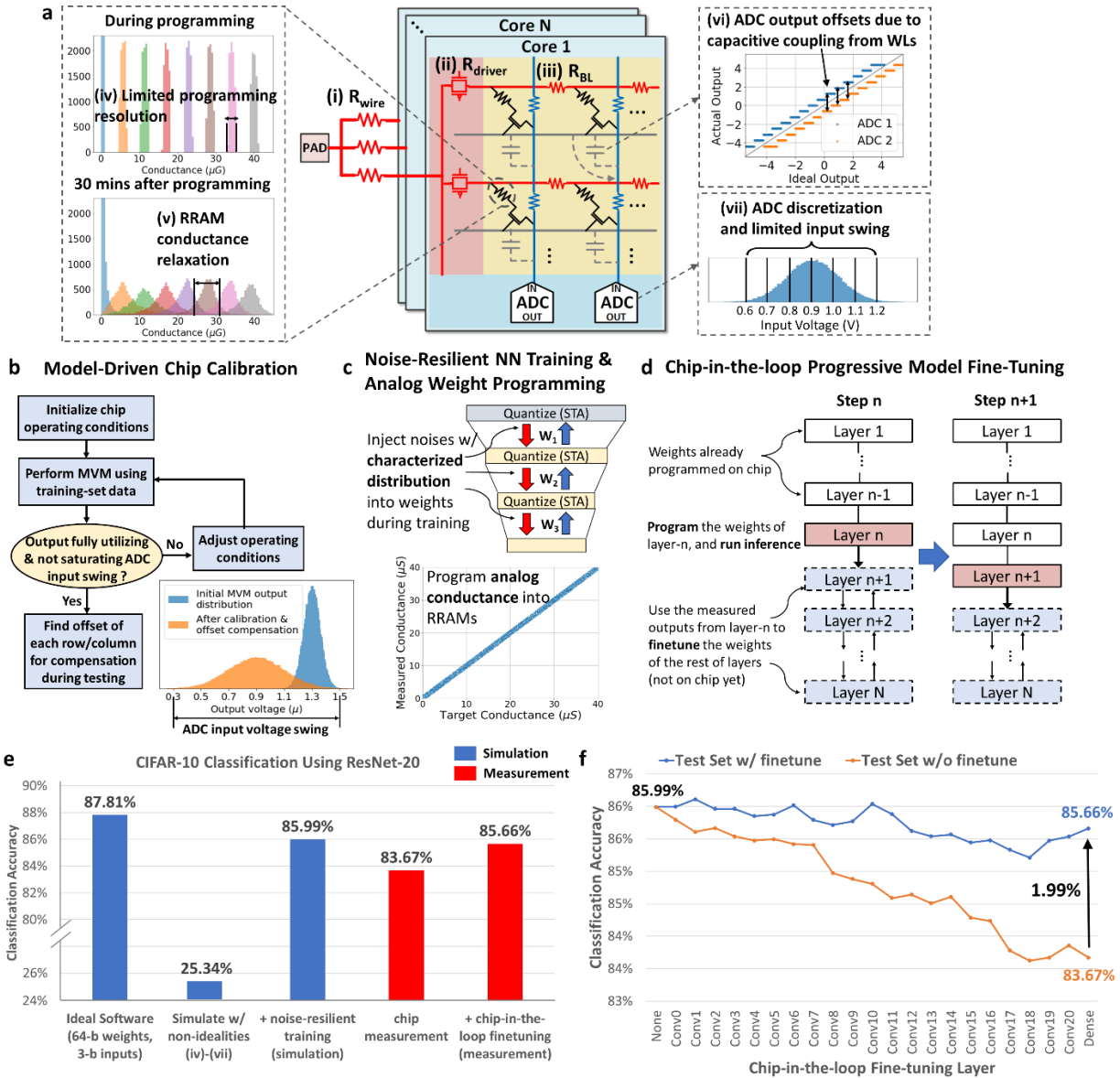

**Fig. 3 | Hardware-algorithm co-optimization techniques to improve NeuRRAM inference accuracy.**
**a**, Various device and circuit non-idealities (labeled (i) to (vii)) of in-memory MVM. **b**, Model-driven chip calibration technique to search for optimal chip operating conditions. **c**, Noise-resilient NN training technique to train the model with injected noise whose distribution is obtained from hardware characterization. The trained weights are programmed to the continuous analog conductance of RRAMs without quantization as shown by the continuous diagonal band at the bottom. **d**, Chip-in-the-loop progressive fine-tuning technique: weights are progressively mapped onto the chip one layer at a time. The hardware measured outputs from layer *n* are used as inputs to finetune the remaining layers *n*+1 to *N*. **e**, Simulated (blue) and measured (red) CIFAR-10 test-set classification accuracy showing the efficacy of the proposed techniques. **f**, CIFAR-10 classification accuracy at various time steps of chip-in-the-loop fine-tuning. From left to right, each data point represents a new layer programmed onto the chip. The accuracy at a layer is evaluated by using the hardware measured outputs from that layer as inputs to the remaining layers that are simulated in software. The two curves compare the test-set inference accuracy with and without applying fine-tuning during training.

| Application | Dataset | Model Architecture | Dataflow Type | Activation Precision | # Parameters |
|---|---|---|---|---|---|
| Image Classification | CIFAR-10 | ResNet-20 (CNN) | Forward | 3-b unsigned (1st layer 4-b unsigned) | 274K |
| | MNIST | 7-Layer CNN | Forward | 3-b unsigned | 23K |
| Voice Recognition | Google Voice Command | 4 parallel LSTM cells | Recurrent + Forward | 4-b signed | 281K |
| Image Recovery | MNIST | RBM | Forward + Backward | Visible: 3-b unsigned. Hidden: binary | 96K |

**Table 1 |** Summary of AI applications and models demonstrated on NeuRRAM

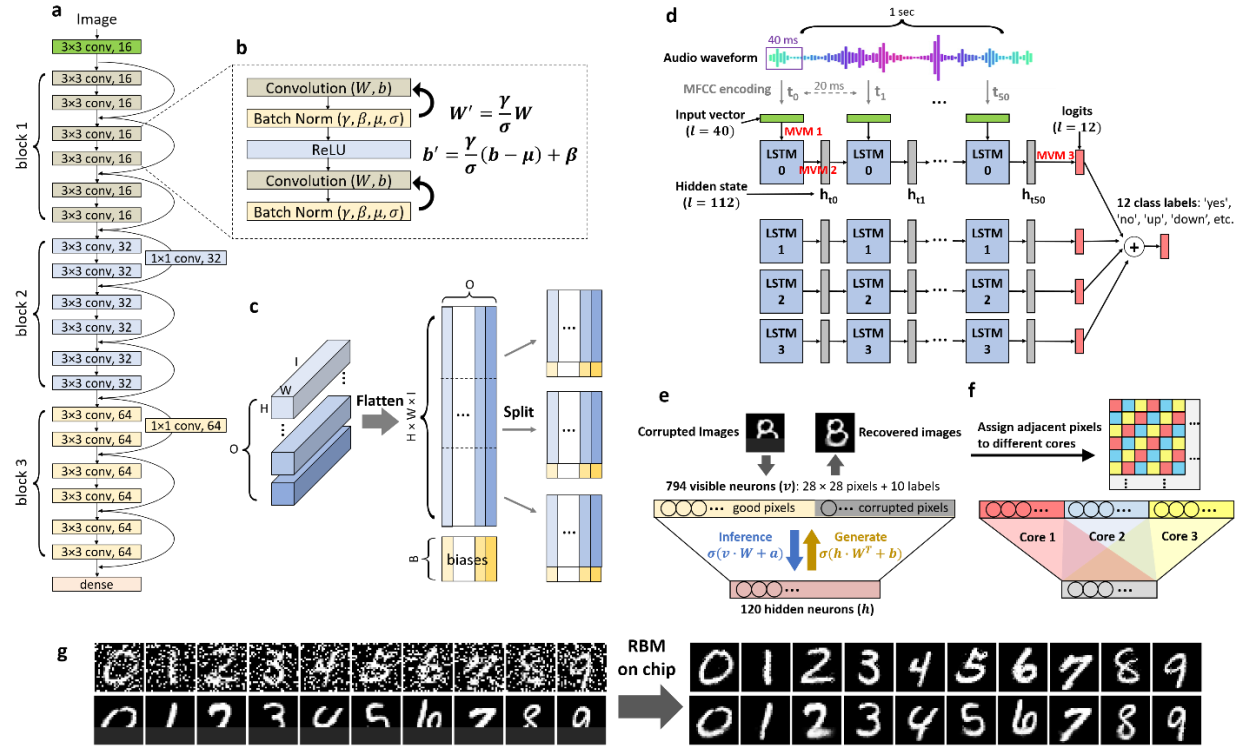

**Fig. 4 | Implementation of various machine learning models.** The implementation details of the models are described in Methods. **a**, Architecture of ResNet-20 for CIFAR-10 classification. **b**, The batch normalization parameters are merged into convolutional layer weights and biases before mapping weights on chip. **c**, Illustration of the process to map the weights of a 4D convolutional layer to NeuRRAM multi-core. **d,** Architecture of the LSTM model used for Google speech command recognition. The model contains 4 parallel LSTM cells and makes predictions based on the sum of outputs from the 4 cells. **e,** Architecture of the RBM model used for MNIST image recovery. During inference, MVMs are performed back-and-forth between visible and hidden neurons. **f,** Process to map RBM on NeuRRAM multi-core. Adjacent pixels are assigned to different cores to equalize the MVM output dynamic range of different cores. **g,** Image recovery on (top) noisy images and (bottom) partially occluded images measured on NeuRRAM.

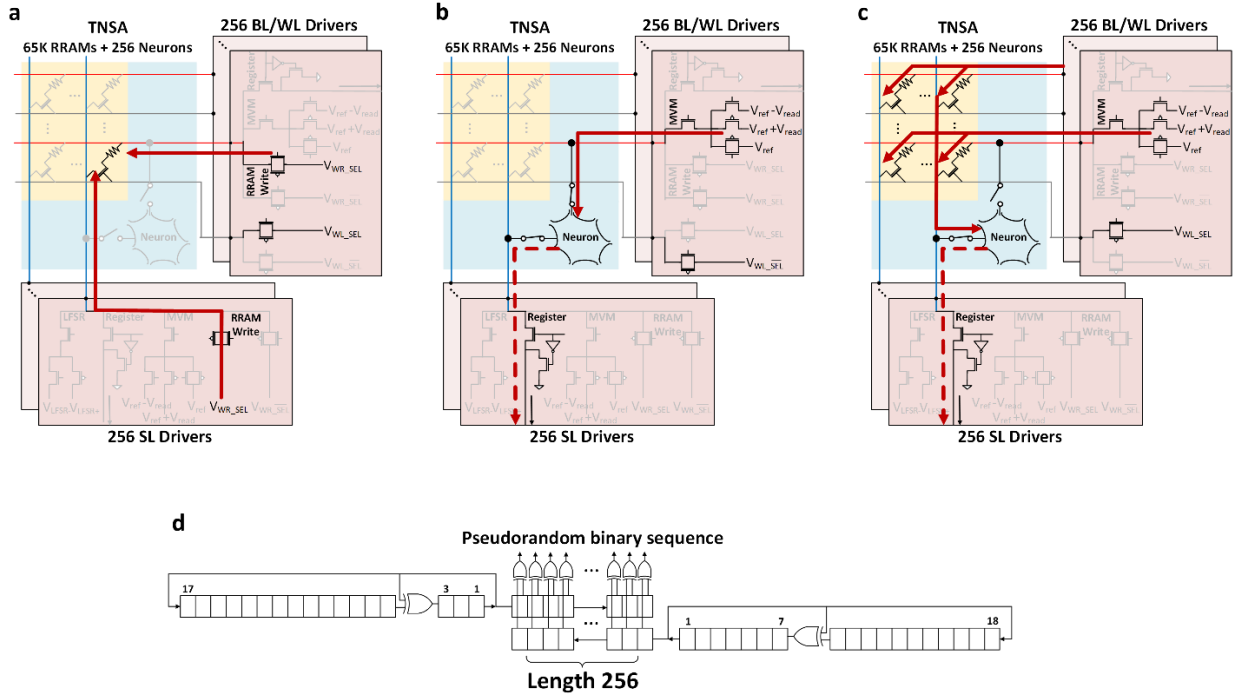

**Extended Data Fig. 1 | Peripheral driver circuits for TNSA and chip operating modes. a**, driver circuits' configuration under the weight-programming mode. **b**, under the neuron testing mode. **c**, under the MVM mode. **d**, circuit diagram of the two counter-propagating LFSR chains XORed to generate pseudorandom sequences for probabilistic sampling.

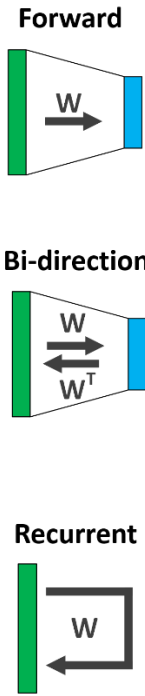
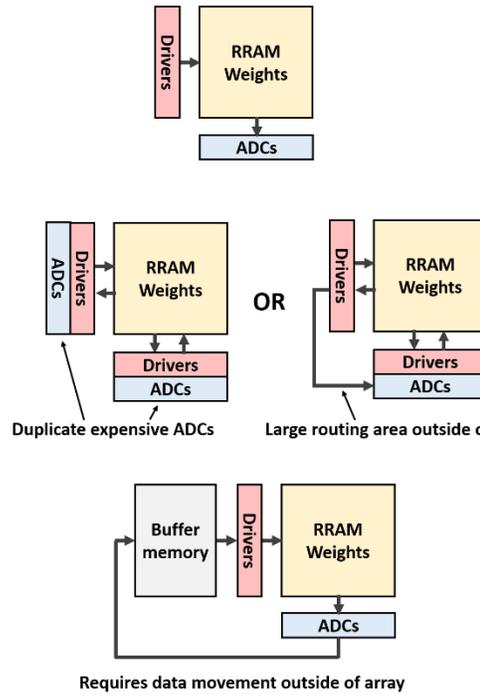
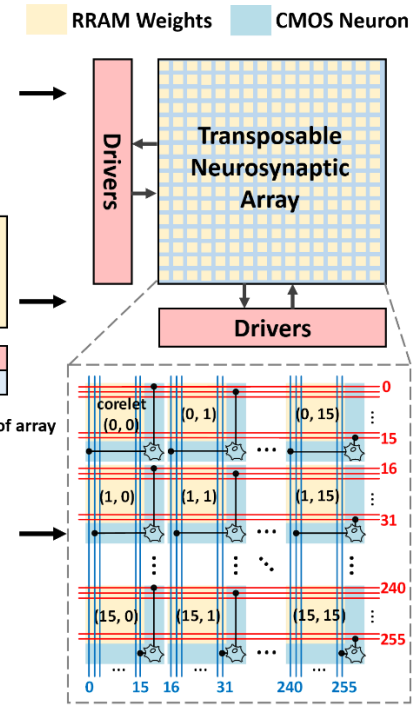

**Extended Data Fig. 2 | Various MVM dataflow directions and compute-in-memory (CIM) implementations. Left,** various MVM dataflow directions commonly seen in different AI models. **Middle,** conventional CIM implementation of various dataflow directions. Conventional designs typically locate all peripheral circuits such as ADCs outside of RRAM array. The resulting implementations of bi-directional and recurrent MVMs incur overheads in area, latency and energy. **Right,** the Transposable Neurosynaptic Array (TNSA) interleaves RRAM weights and CMOS neurons implementing ADCs and activation functions across the array and supports diverse MVM directions with minimal overhead.

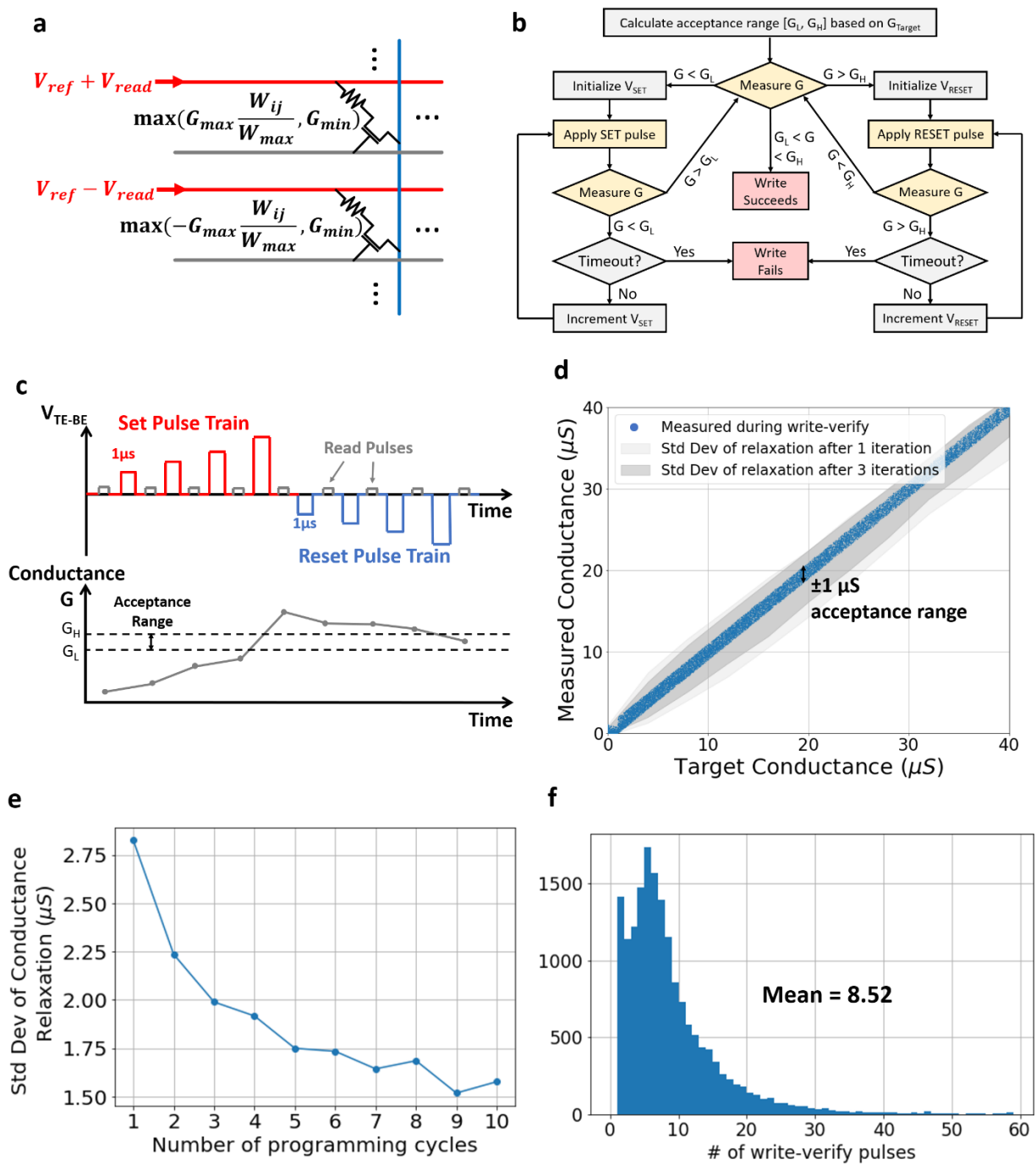

**Extended Data Fig. 3 | Iterative write-verify RRAM programming technique. a**, Weights of neural networks are encoded as the differential conductance between two RRAM cells on adjacent rows. **b**, Flowchart of the incremental-pulse write-verify technique to program RRAMs into target analog conductance range. **c**, An example sequence of the write-verify programming. **d**, The RRAM conductance distribution measured during and after the write-verify programming. Each blue dot represents one RRAM cell. The gray shades show that the RRAM conductance relaxation cause the distribution to spread out from the target values. The darker shade shows that the iterative programming helps narrow the distribution. **e,** The standard deviation of conductance relaxation decreases with increasing iterative programming cycles. **f,** Distribution of the number of SET/RESET pulses needed to reach the acceptance range.

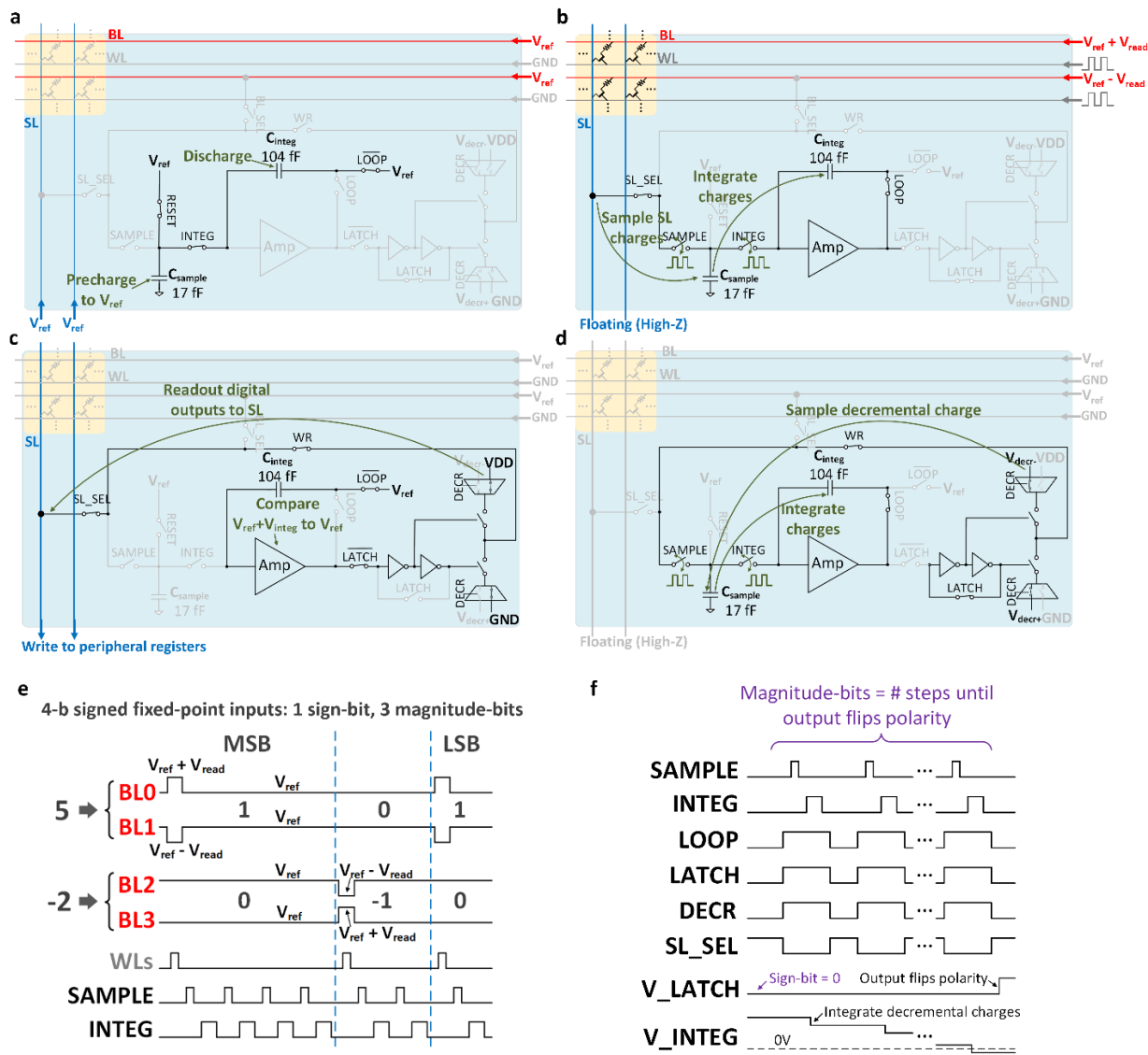

**Extended Data Fig. 4 | Neuron and RRAM array operations to perform MVM and ADC with multi-bit inputs and outputs. a**, Initialization phase, precharging the sampling capacitor and output wires (SLs), and discharging the integration capacitor. **b**, Input phase, pulsing the WLs. Upon settling, the SL voltage is sampled onto $C_{sample}$, and its charge is accumulated onto $C_{integ}$. Different combinations of WL, sampling and integration clocks enable configurable input bit-precision. **c**, Output phase to generate the sign-bit. The amplifier is turned into comparator mode to determine the polarity of the integrated voltage. Comparator outputs are written out of the neuron through the outer feedback loop. **d**, Output phase to generate the magnitude-bits. During each step, a small amount of charge is subtracted from $C_{integ}$ through the outer feedback loop. **e,** Sample waveforms to perform MVM with 4-bit signed inputs. The BLs are driven to voltages corresponding to each bit; WLs are pulsed 3 times, and the SAMPLE and INTEG switches are pulsed $2^{4-1} – 1 = 7$ times. **f,** Sample waveforms to perform MVM with multi-bit outputs. The comparison and charge decrement steps are repeated until the comparator output flips polarity. The number of steps correspond to the magnitude-bits of MVM outputs.

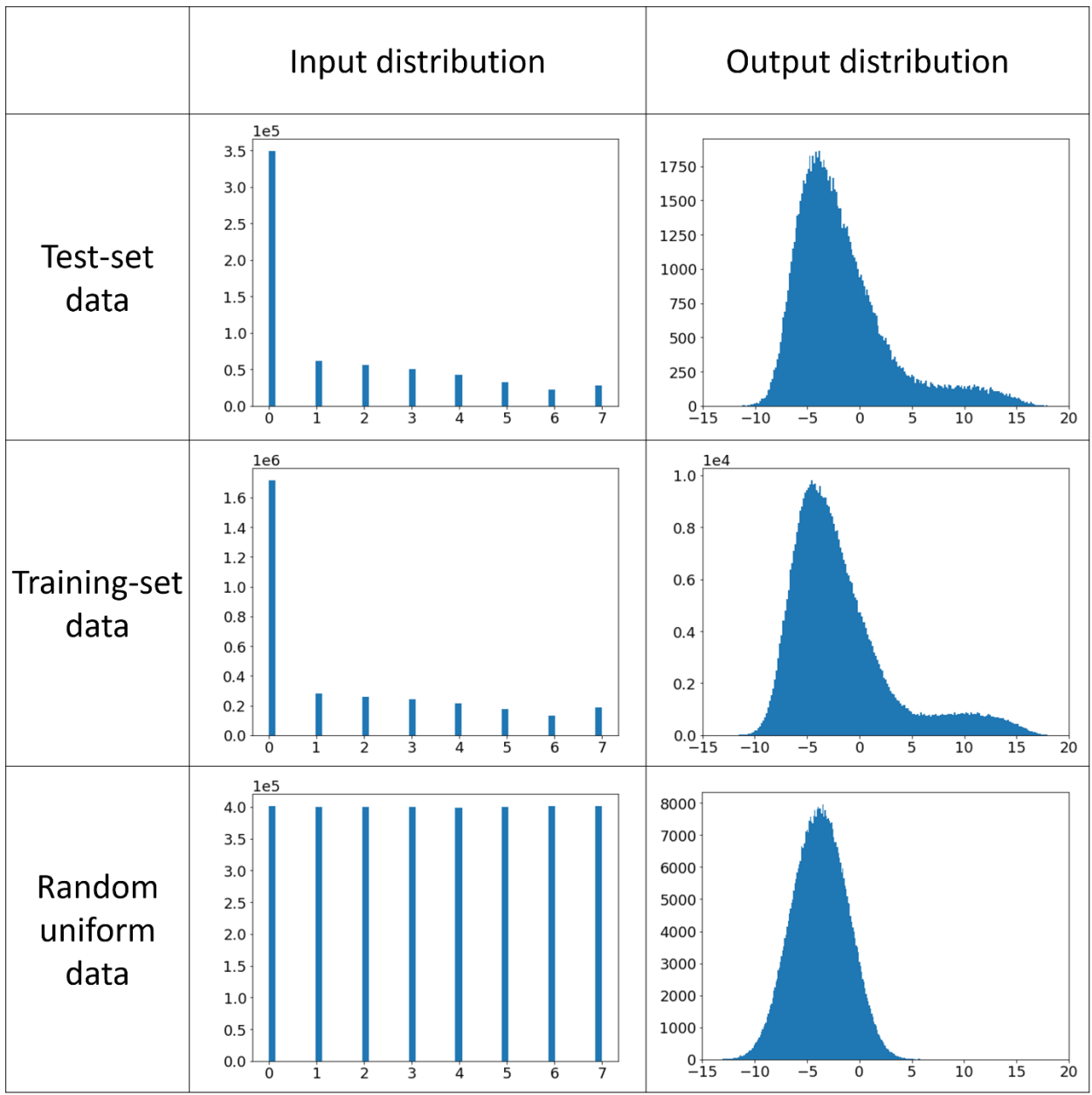

**Extended Data Fig. 5 | Data distribution with and without model-driven chip calibration. Left**, Distribution of inputs to the final fully-connected layer of ResNet-20 when the inputs are generated from CIFAR-10 test-set data, training-set data, and random uniform data. **Right**, Distribution of outputs from the final fully-connected layer of ResNet-20. The test-set and training-set have similar distributions while random uniform data produces a markedly different output distribution. To ensure that the MVM output voltage dynamic range during testing is calibrated to occupy the full ADC input swing, the calibration data should come from training-set data that closely resembles the test-set data.

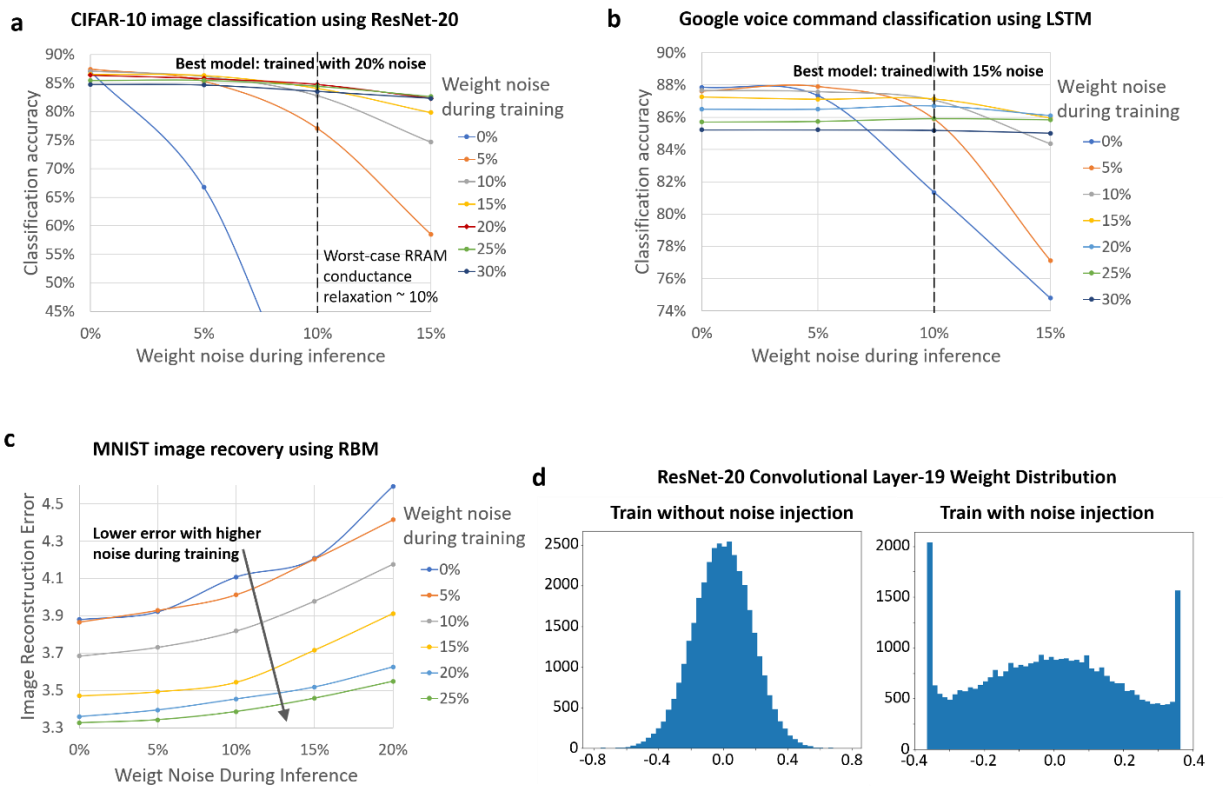

**Extended Data Fig. 6 | Noise resilient training of CNNs, LSTMs and RBMs. a**, Change in CIFAR-10 validation-set classification accuracy under different weight noise levels during inference. Noise is represented as fraction of the maximum absolute value of weights. Different curves represent models trained at different levels of noise injection. **b**, Change in voice command recognition accuracy with weight noise levels. **c**, Change in MNIST image reconstruction error with weight noise levels. **d**, Differences in weight distributions when trained without and with noise injection.

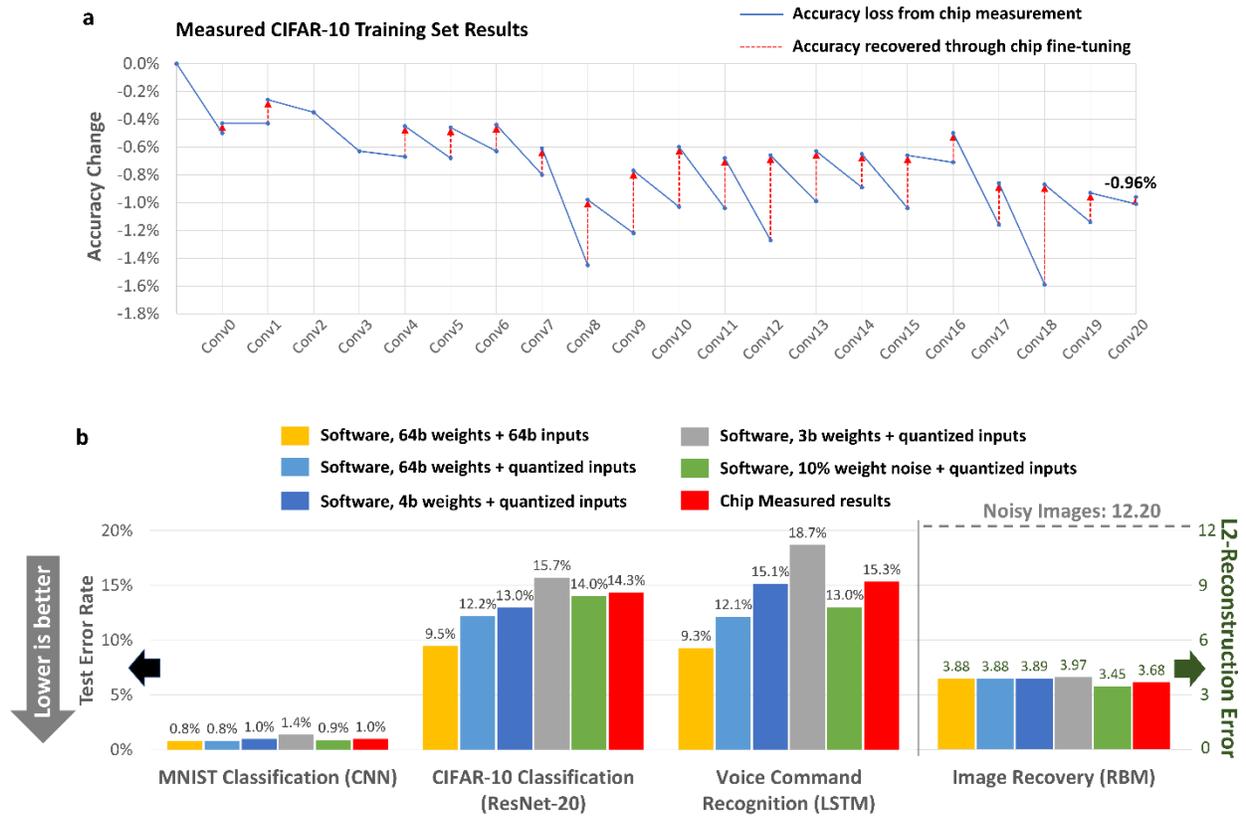

**Extended Data Fig. 7 | Measured chip inference performance. a,** CIFAR-10 training-set accuracy loss due to hardware non-idealities, and accuracy recovery at each step of the chip-in-the-loop progressive fine-tuning. From left to right, each data point represents a new layer programmed onto the chip. The blue solid lines represent the accuracy loss measured when performing inference of that layer on chip. The red dotted lines represent the measured recovery in accuracy through fine-tuning of that layer, retrained along with the subsequent layers of the network using chip measured training data from that layer. **b,** Comparison of inference errors on various applications between chip measured results (red bars) and software models with different configurations of input quantization, weight quantization, and weight noise injection.

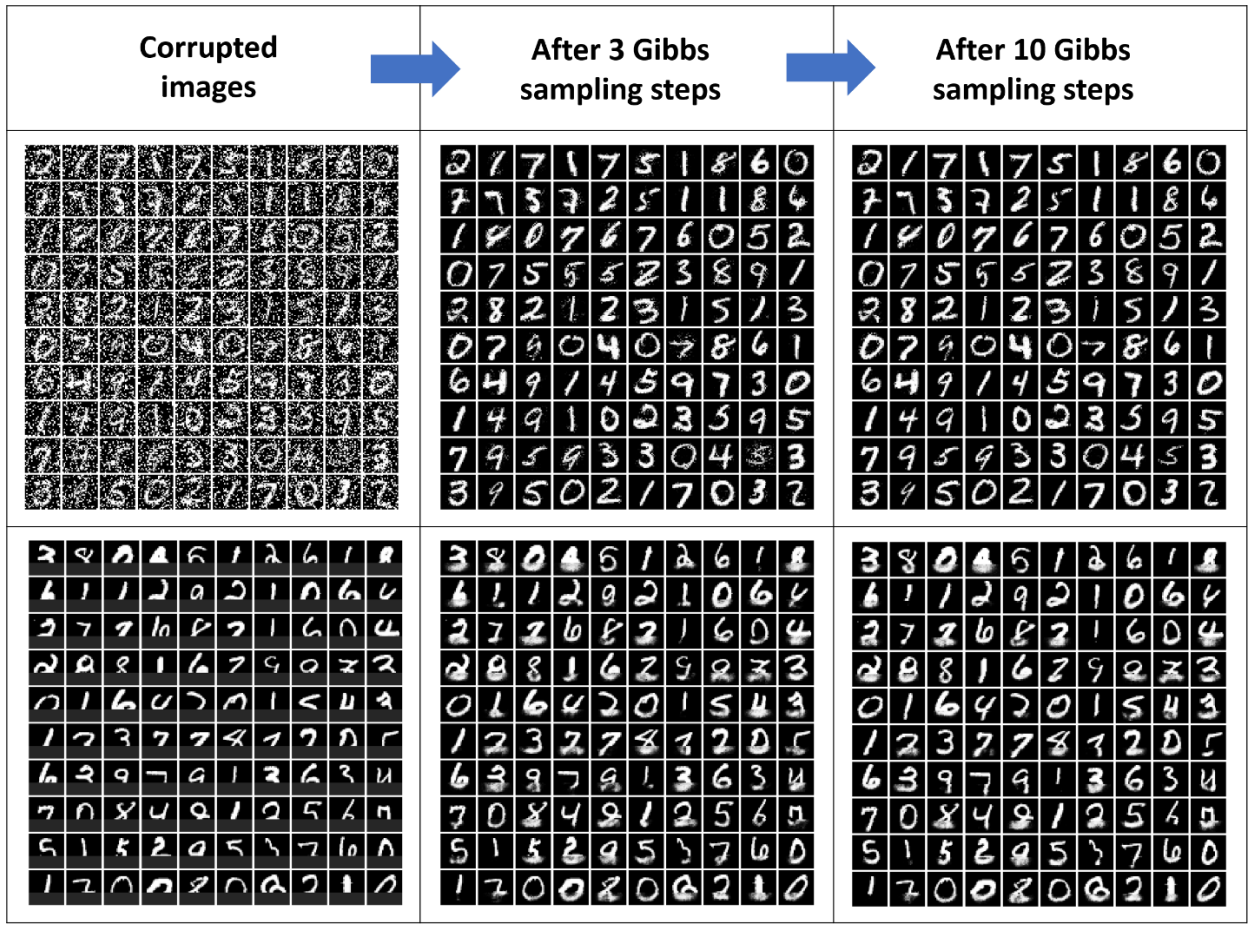

**Extended Data Fig. 8 | Chip measured image recovery using RBM. Top:** Recovery of MNIST test-set images with random 20% of pixels flipped to complementary intensity. **Bottom:** Recovery of MNIST test-set images with bottom 1/3 of pixels occluded.

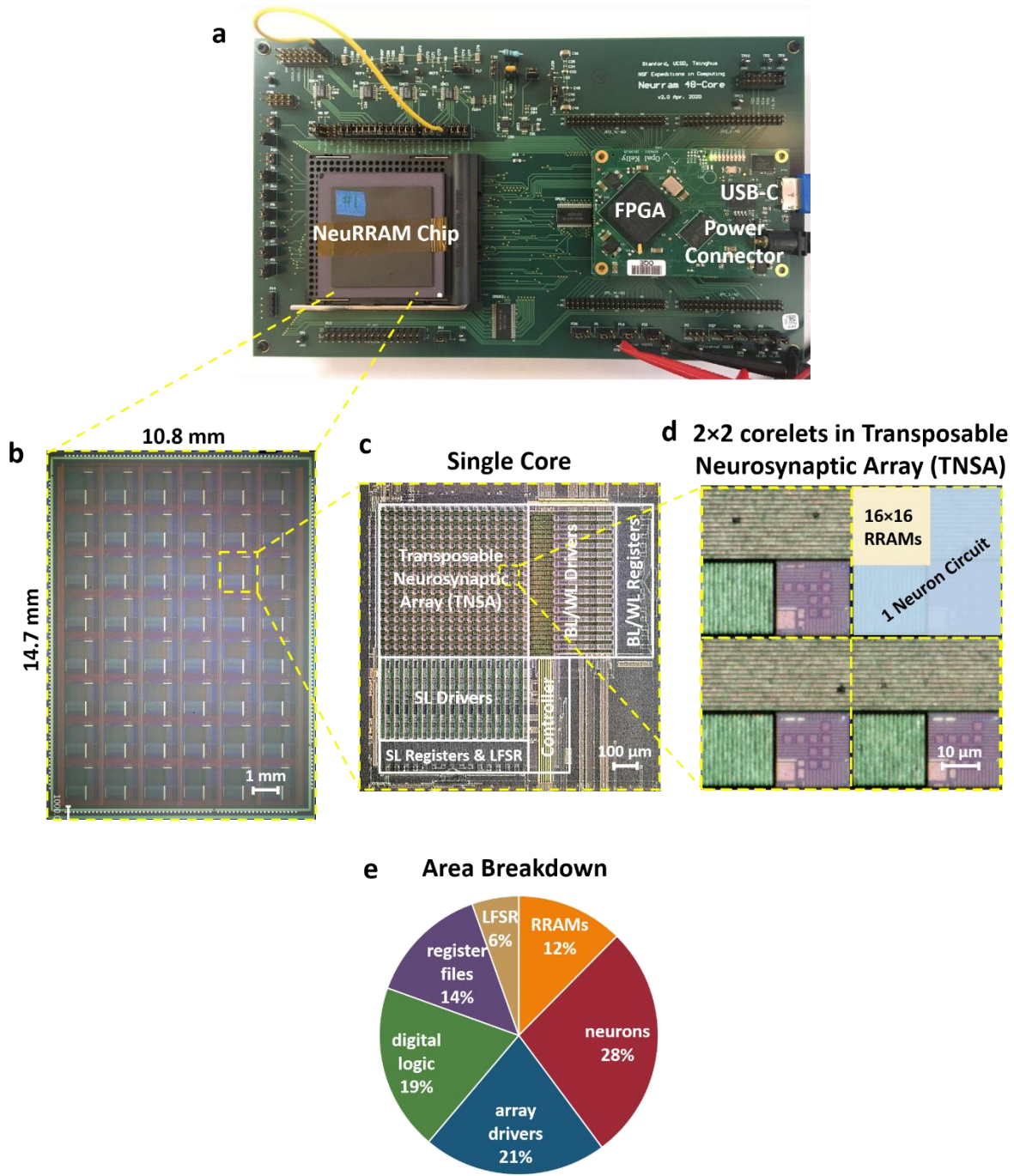

**Extended Data Fig. 9 | NeuRRAM test system and micrographs at various scales. a**, Measurement board that connects the NeuRRAM chip (left) to a field-programmable gate array (FPGA, right) for input/output and clock waveform control. The board houses all the components necessary to power, operate and measure the chip. No external lab equipment is needed for the chip operations. **b**, Micrograph of the 48-core NeuRRAM chip. **c,** Zoomed-in micrograph of a single compute-in-memory core. **d,** Zoomed-in micrograph of 2×2 corelets within the TNSA. **e,** Chip area breakdown.

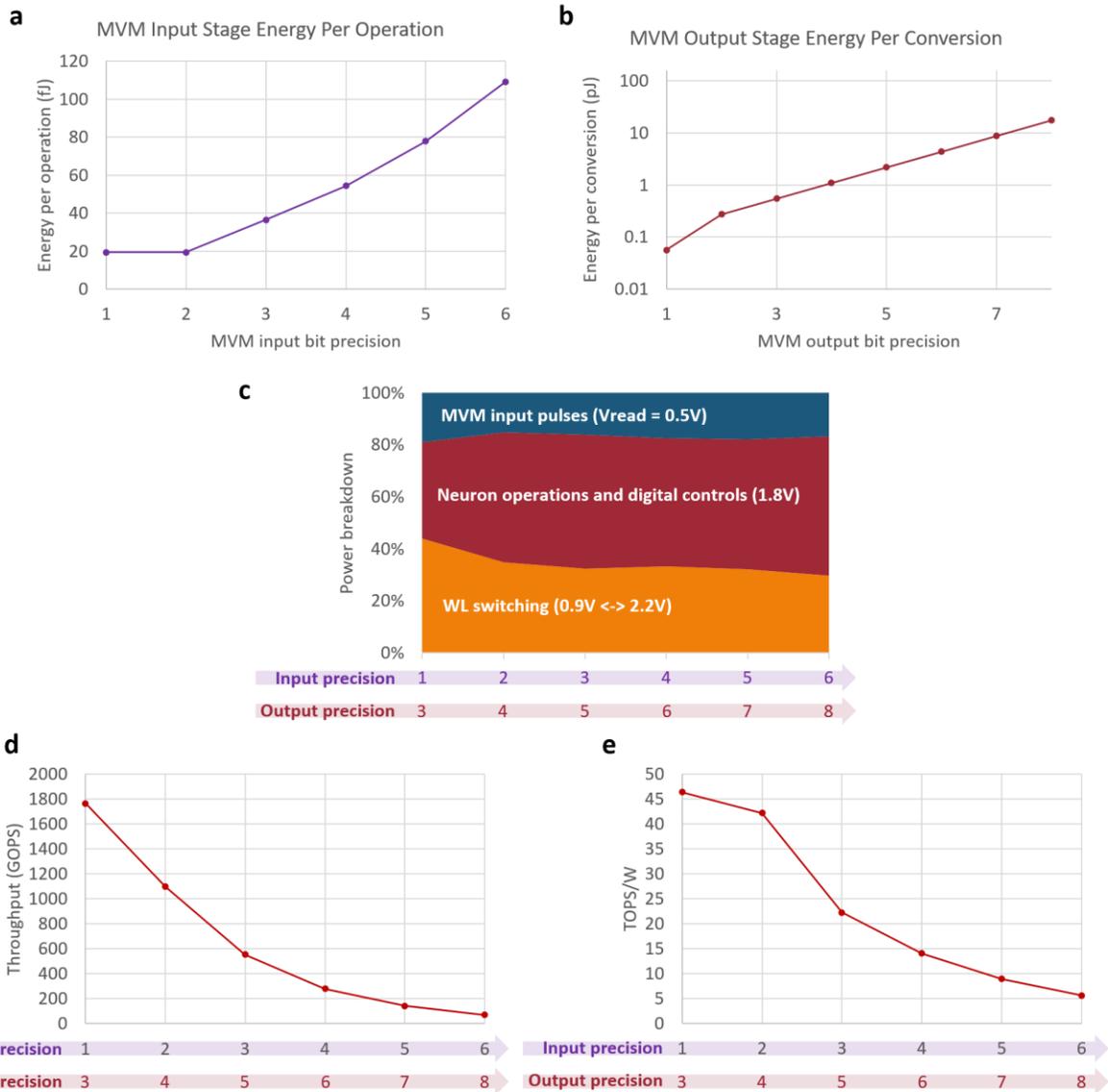

**Extended Data Fig. 10 | Power consumption and throughput measurement results. a**, Measured energy consumption per operation during the MVM input stage, where one multiply-accumulate (MAC) counts as two operations. **b**, Measured energy consumption per analog-to-digital conversion during the MVM output stage. **c,** Power consumption breakdown during the MVM input stage. **d,** Peak computational throughput (in giga-operations per second) measured at various MVM input and output bit-precisions. **e,** Throughput-power efficiency (in tera-operations per watt) measured at various bit-precisions.